\documentclass[12pt,english,a4paper]{article}

\pdfoutput=1

\usepackage{amsmath,amsfonts,amssymb,babel,slashed,color,empheq,tensor}

\usepackage{jheparxiv}
\usepackage[utf8]{inputenc}
\usepackage{amsmath}
\usepackage{amsfonts}
\usepackage{amssymb}
\usepackage{latexsym}
\usepackage{mathrsfs}
\usepackage{braket}		
\usepackage{setspace}
\usepackage{graphicx}
\usepackage{color}
\usepackage{xcolor}
\usepackage{slashed}
\usepackage{twistor}
\usepackage[all]{xy}
\usepackage{tikz-cd}
\usepackage{float}
\usepackage{mathtools}
\usepackage{tikz}

\makeatletter

\newcommand{\bep}{\begin{picture}}
\newcommand{\eep}{\end{picture}}

\newcounter{YoungHeight}\newcounter{YoungWidth}

\newcounter{Mul1}\newcounter{Mul2}\newcounter{Mul3}\newcounter{Mul4}
\newcounter{A0}\newcounter{A1}\newcounter{A2}
\newcounter{B3}
\newcounter{C3}\newcounter{C4}
\newcounter{D1}\newcounter{D2}\newcounter{D3}
\newcounter{T0}\newcounter{T1}

\newlength{\txtHShift}

\newlength{\txtWidth}

\newcommand{\ib}{\boldsymbol{\mathtt{I}}}
\newcommand{\jb}{\boldsymbol{\mathtt{J}}}

\newcommand{\hatkappa}{\hat{\kappa}}
\newcommand{\cY}{\hat{\mathcal{Y}}}
\newcommand{\ttb}{\underline{\mathtt{t}}}
\newcommand{\cS}{\widetilde{\mathcal{S}}}
\newcommand{\Ibold}{\boldsymbol{I}}
\newcommand{\Jbold}{\boldsymbol{J}}

\newcommand{\BG}{\mathtt{BG}}

\newcommand{\tp}{\mathsf{p}}

\usepackage{graphicx}

\newcommand{\nn}{\nonumber}
\newcommand{\tJ}{\mathtt{J}}

 \def\one{\mbox{1 \kern-.59em {\rm l}}}

\newcommand{\sa}{\mathsf{a}}

\newcommand{\eps}{\epsilon}

\newcommand{\msf}[1]{\mathsf{#1}}

\newcommand{\sF}{\mathsf{F}}

\newcommand{\msu}{\mathfrak{su}}

\def\l{\lambda} \def\L{\Lambda}  

\makeatother

\begin{document}

 \begin{flushright}
  UWThPh-2023-27 
 \end{flushright}
 \vspace{-10mm}
 
\title{\Large Interactions in the IKKT matrix model on covariant quantum spacetime}

\author[]{Harold C. Steinacker}
\affiliation[]{Department of Physics, University of Vienna, \\
Boltzmanngasse 5, A-1090 Vienna, Austria}
\emailAdd{harold.steinacker@univie.ac.at}

\author[]{\& Tung Tran}

\emailAdd{tung.tran@univie.ac.at}

\abstract{We study the interactions of the higher-spin gauge theory arising from the IKKT matrix model on a covariant quantum FLRW quantum space-time $\cM^{1,3}_{\tJ}$, denoted as HS-IKKT. In particular, we elaborate some of the vertices 
and observe that they are not manifestly Lorentz invariant in the unitary formulation. 
We argue that Lorentz invariance of HS-IKKT can be recovered since Lorentz transformations are part of the 
gauge invariance in the covariant formulation. 
This statement is verified for some vertices with lowest number of derivatives. The lowest-derivative sector of this theory is expected to be governed by an ``almost"-Lorentz-invariant Yang-Mills theory coupled to emergent gravity. 
}

\maketitle

\section{Introduction}
The idea that gravity can emerge from 
suitable non-gravitational or ``pre-gravity'' models dates back to the late 60's \cite{Sakharov:1967pk,Visser:2002ew}. Such a mechanism was recently exhibited in the IKKT matrix model \cite{Ishibashi:1996xs}, which can be viewed as a constructive description of type IIB superstring theory.\footnote{Roughly speaking, the IKKT matrix model can be understood as a superstring theory with open strings propagating on a $D$-brane $\cM$ embedded in $\R^{1,9}$ target space in the large $N$ limit. See e.g. \cite{Steinacker:2019fcb} for a review.} In particular, it was shown that a version of modified GR can arise from quantum effects of the IKKT matrix model on a suitable type of background \cite{Steinacker:2021yxt,Steinacker:2023myp}. The mechanism requires $3+1$ dimensional quantized spacetime branes and certain compact fuzzy extra dimensions without compactifying target space, and thereby
avoids the problem of vacuum selection in the string landscape. For related progress on the non-perturbative selection of spacetime using a numerical approach, see e.g. \cite{Anagnostopoulos:2022dak,Nishimura:2019qal} and references therein.

Due to the presence of a $B$ field on the background brane, the  theories induced by the IKKT matrix model will typically be Lorentz-violating. Intriguingly,  such  explicit breaking of Lorentz invariance can be reduced by considering the IKKT matrix model on a suitable type of covariant quantum background brane $\cM\xhookrightarrow{}\R^{1,9}$ such as the fuzzy 4-sphere $S_{\tJ}^4$, the fuzzy 4-hyperboloid $H_{\tJ}^4$, or an analogous Lorentzian space $\cM^{1,3}_\tJ$   considered below. On such a background, the matrix model naturally induces a higher-spin gauge theory 
in terms of higher-spin valued fluctuations, see e.g. \cite{Sperling:2018xrm,Steinacker:2023zrb}. Unlike the standard higher-spin gravity  in $d\geq 4$ \cite{Fradkin:1986ka} with a spectrum of infinitely many fields with spin $s=0,1,\ldots,\infty$, the higher-spin gauge theory induced by the IKKT matrix model (HS-IKKT) typically leads to a truncated tower of higher-spin modes in 4 dimensions. Moreover, the number of physical degrees of freedom for a higher-spin gauge field with \emph{internal} spin-$s$ in HS-IKKT theory is $2(2s+1)$. These on-shell physical dof. are interpreted as the ``would-be-massive'' modes \cite{Steinacker:2019awe},
because there are ``extreme IR" 
mass terms set by the cosmic curvature scale associated to them. Nevertheless, the theory admits a very large gauge group, which can be interpreted as a higher-spin extension of volume-preserving diffeos \cite{Sperling:2019xar}.

In this work, we study the HS-IKKT theory on the $SO(1,3)$-covariant cosmological background called $\cM^{1,3}_\tJ$, with $\tJ$ being the cutoff of the internal spins. This background is obtained by a suitable projection of the fuzzy 4-hyperboloid $H^4_\tJ$. Locally, $\cM^{1,3}_\tJ$ is isomorphic to $S^2_{\tJ}\times \cM^{1,3}$, where $S^2_{\tJ}$ is the internal 2-sphere at each point of an FLRW spacetime with $k=-1$ denoted as $\cM^{1,3}$. This internal 2-sphere is responsible for inducing higher-spin structures of the model as we will elaborate in the main text. Due to the Yang-Mills structure of the theory and the space-like structure of $S^2_{\tJ}$, the resulting gauge theory can be shown to be free of ghosts (negative norm states) \cite{Steinacker:2019awe}.



Since Lorentzian HS-IKKT is not massless in the strict sense, 
the standard implications of no-go theorems such as \cite{Weinberg:1964ew,Coleman:1967ad} do not apply to this theory. However, it is still a gauge theory which is ``almost" massless, 
and hence it is natural to ask how  the standard arguments realted to soft theorems or factors resulting from gauge invariance of the $S$-matrix can be circumvented\footnote{See \cite{Steinacker:2023zrb} for recent work on tree-level scattering amplitudes of massless higher-spin modes within the Yang-Mills sector of Euclidean HS-IKKT.}.
In this work, we 
elaborate some vertices of the Lorentzian theory explicitly and study their physical significance. 

To study the interactions of HS-IKKT, we will focus on the locally $4d$ regime, where a patch of $\cM^{1,3}$ can be treated effectively as a Minkowski spacetime. We observe that gauge invariance of the $S$-matrix in the soft limit is trivially satisfied, incorporating the Lorentz-violating structure of the cubic vertices.
Hence, there are no non-trivial constraints in terms of external momentum in the soft limit for which we could extract soft theorems in the sense of \cite{Weinberg:1964ew}. Although this may seem counter-intuitive, it is a direct consequence of the manifest but non-standard gauge invariance of the model, which includes volume-preserving diffeomorphisms. Note that Lorentz invariance is expected to be recovered as part of the gauge symmetry.

We also observe that the cubic vertices for the lowest spin components of the fields vanish. 
Nevertheless, the theory is not free: non-trivial cubic vertices involving higher-spin modes do arise, which are suppressed in the late-time, almost-flat regime of the theory. Furthermore, the quartic vertices of the theory are non-trivial even with the lowest spin components of the higher-spin fields. In particular, these vertices are 
Lorentz-violating, as they involve the explicit time-like vector field 
defined by the FLRW background geometry.
We show further that  all cubic vertices are suppressed by a factor of $L^2_{\rm NC}\,e^{-\frac{3\tau}{2}}$ in the late time regime,\footnote{Non-trivial gravitational interactions do arise in the one-loop effective action, see e.g. \cite{Steinacker:2021yxt,Steinacker:2023myp}.} while quartic vertices are suppressed by a factor of $L^4_{\rm NC}\,e^{-3\tau}$. Here, $\tau$ parametrizes the cosmic time of the FLRW background, and $L_{\rm NC}$ is a non-commutative length scale. As a simple exercise, we compute some 3-amplitudes of HS-IKKT with the obtained vertices explicitly. We also spell out the fusion rules, and study the simplest 4-point tree-level amplitude.

Finally, we observe that at least some of these apparently Lorentz-violating interactions can in fact be rewritten in a manifestly covariant form, consistent with the expectation that Lorentz invariance is recovered from gauge invariance. This suggests that Lorentz invariance should be understood here in a  non-standard way as part of the gauge symmetry.

This paper is organized as follows. Section \ref{sec:2} provides a brief review of the (HS-)IKKT matrix model on a fuzzy $SO(1,3)$-invariant spacetime called $\cM^{1,3}_{\tJ}$. Section \ref{sec:3} derives some cubic and quartic vertices between (higher-spin) fields of the model. The soft limit of higher-spin scatterings associated with the cubic vertices is also discussed. Some 3-point amplitudes are computed explicitly, and the simplest 4-point amplitude is also studied. Finally, Section \ref{sec:discussion} summarizes the results and discusses future direction.

\section{Review of the higher-spin IKKT model}\label{sec:2}

We consider the semi-classical form of the $SO(1,9)$-invariant IKKT matrix model in the
large $N$ limit, given by the action
\begin{align}\label{SO(1,9)action}
    S= -\int_{\cM} \,\Big(\{t^{\Ibold},t^{\Jbold}\}\{t_{\Ibold},t_{\Jbold}\}+2m^2t_{\Ibold}t^{\Ibold}+\ldots\Big)\,, \qquad \Ibold=0,1,\ldots,9\,.
\end{align}
Here $t^{\Ibold}:=\langle \zeta|T^{\Ibold}|\zeta\rangle$ play the role of embedding functions of some underlying quantized brane $\cM$ in $\R^{1,9}$, defined by the expectation values of $N\times N$ Hermitian background matrices $T^{\Ibold}$ w.r.t. suitable
localized quasi-coherent states $|\zeta\rangle\in \cH$ of some Hilbert space $\cH$. The Poisson brackets arise from commutators of the matrix model in the semi-classical regime. In this language, the action is invariant under infinitesimal gauge transformations $t^{\Ibold} \to t^{\Ibold} + \{t^{\Ibold},\L\}$ for $\Lambda$ a gauge parameter.\footnote{This geometrical approach to the matrix model is reviewed e.g. in \cite{Steinacker:2019fcb}.}
The ellipses in \eqref{SO(1,9)action} stand for the fermionic part, and a mass term $m^2$ is added to set the scale for the theory\footnote{This mass term is not essential, but it allows stabilizing the background at the classical level. The IR mass terms for the higher-spin modes arise independently from the Box operator.}.

For our purpose of studying
tree-level scattering amplitudes, we will consider the interactions between the higher-spin valued scalar fields transforming under $SO(6)$, and higher-spin gauge fields. It is thus sufficient to consider only the bosonic part of the model. 


\paragraph{Covariant quantum spacetime.} 

In this work, we are interested in studying scattering amplitudes of HS-IKKT theory in the local $4d$ regime of a covariant quantum spacetime\footnote{For example, CERN is located on a local patch of our curved universe
.}  called $\cM^{1,3}_{\tJ}$. 
Here, $\cM^{1,3}_{\tJ}$ is a $S^2_{\tJ}$ bundle over a $k=-1$ FLRW cosmological spacetime $\cM^{1,3}$. This can be viewed as a $3+1$ dimensional degenerate embedding of the non-compact complex projective space $\P^{1,2}$ in the target space. 
The internal 2-sphere $S^2_{\tJ}$ attached to each point on $\cM^{1,3}$ encodes higher-spin structures of the model up to some cutoff $\tJ$. 
 
Let $\ttb^{\Ibold}$ be the  background of the model in the semi-classical limit. Then, we consider the following fluctuations around $\ttb^{\Ibold}$ where
\begin{align}\label{BG:t}
    t^{\Ibold}=\binom{\ttb^{\dot\mu}}{0}+\binom{\sa^{\dot\mu}}{\phi^{\ib}}\,,\qquad \dot\mu=0,1,2,3\,,\qquad \ib=4,\ldots,9\,.
\end{align}
Here, the tangential fluctuations $\sa^{\dot\mu}$ play the role of (non-commutative, $\hs$-valued) gauge fields, and the 6 transversal fluctuations $\phi^{\ib}$ can be viewed as ($\hs$-valued) scalar fields. 
With this organization, the action \eqref{SO(1,9)action} can be written as \cite{Steinacker:2023zrb}
\begin{align}\label{FLRWaction}
    S=-\int &\mho\,\Big(\frac 12 \{\ttb^{\dot\mu},\sa^{\dot\nu}\}\{\ttb_{\dot\mu},\sa_{\dot\nu}\}+\{\ttb_{\dot\mu},\ttb_{\dot\nu}\}\{\sa^{\dot\mu},\sa^{\dot\nu}\}+\frac{1}{2}\{\ttb^{\dot\mu},\sa^{\dot\mu}\}^2+\frac{1}{2}\{\ttb^{\dot\mu},\phi^{\ib}\}\{\ttb_{\dot\mu},\phi_{\ib}\}\nn\\
    &+\{\ttb^{\dot\mu},\sa^{\dot\nu}\}\{\sa_{\dot\mu},\sa_{\dot\nu}\}+\{\ttb^{\dot\mu},\phi^{\ib}\}\{\sa_{\dot\mu},\phi_{\ib}\}\nn\\
    &+\frac{1}{4}\{\sa^{\dot\mu},\sa^{\nu}\}\{\sa_{\dot\mu},\sa_{\dot\nu}\}+\frac{1}{2}\{\sa^{\dot\mu},\phi^{\ib}\}\{\sa_{\dot\mu},\phi_{\ib}\}+\frac{1}{4}\{\phi^{\ib},\phi^{\jb}\}\{\phi_{\ib},\phi_{\jb}\}+\ldots\Big)+S_{\BG} \ ,
\end{align}
where $S_{\BG}$ is the background action, which consists of terms that are zeroth or first order in fields. Here
\begin{align}\label{eq:induced-measure}
    \mho:=\mho_0\times \varpi\,,\qquad \mho_0=\frac{R}{y^4}dy^0dy^1dy^2dy^3=(\sinh(\tau))^{-1}d^4y
\end{align}
is the induced (symplectic) measure, and $y^a:=\langle \zeta|Y^a| \zeta\rangle\in \R^{1,4}$ for $a=0,1,2,3,4$ are Cartesian coordinates of $H^4$. This 4-hyperboloid of radius $R$ is defined by $\eta^{ab}y_ay_b=-R^2$, where $\eta^{ab}=\diag(-,+,+,+,+)$.
The $SO(1,3)$-invariant space-time $\cM^{1,3}$ is then obtained as a projection of $H^4$ to $\R^{1,3}$ with $y^4$ removed. We can then describe $\cM^{1,3}$ by the following global hyperbolic coordinates
\begin{align}
    \begin{pmatrix}
     y^0\\
     y^1\\
     y^2\\
     y^3
    \end{pmatrix}=R\cosh(\tau)\begin{pmatrix}
     \cosh(\chi)\\
     \sinh(\chi)\sin(\theta)\cos(\varphi)\\
     \sinh(\chi)\sin(\theta)\sin(\varphi)\\
     \sinh(\chi)\cos(\theta)
    \end{pmatrix}\,.
\end{align}
Here, the time parameter 
$\tau$ defined via $y_4=R\sinh (\tau)$.
The $\varpi$ in \eqref{eq:induced-measure} is the symplectic form on $S^2_{\tJ}$ which is normalized as\footnote{Strictly speaking, we should consider the normalization $\int_{S^2_{\tJ}}\varpi=\tJ$ corresponding to the trace over the Hilbert space $\C^{\tJ}$ associated with $S^2_{\tJ}$. However, this simply results in an overall $\tJ$ factor in the action, which is irrelevant here. This factor is, however, important for studying loop corrections.}
\begin{align}
    \int_{S^2_{\tJ}}\varpi=1\,.
\end{align}
On this spacetime $\cM^{1,3}$, the Poisson brackets between $\ttb^{\dot\mu}$ and $y^{\nu}$ define a frame
\begin{align}\label{eq:PoissonM13}
    E^{\dot\mu\nu} := \{\ttb^{\dot\mu},y^{\nu}\}&=\frac{\eta^{\dot\mu\nu}}{R}y^4=\eta^{\dot\mu\nu}\sinh(\tau)\,,\qquad \eta^{\dot\mu\nu}=\diag(-,+,+,+)\,,
\end{align}
for $\dot\mu,\nu=0,1,2,3$. Note that dotted indices will be referred to as the frame or local Lorentz indices while the undotted ones are spacetime indices.
Using the frame, we can define the auxiliary metric $\gamma^{\mu\nu}$ and effective metric $G^{\mu\nu}$ as
\begin{align}
    \gamma^{\mu\nu}=E^{\dot\kappa\mu}E_{\dot\kappa}{}^{\nu}=\sinh^2(\tau)\eta^{\mu\nu}\,,\qquad G^{\mu\nu}=\rho^2\gamma^{\mu\nu}\,,\qquad \rho^2=\sinh^3(\tau)\,,
\end{align}
where $\eta^{\mu\nu}=\diag(-,+,+,+)$. Here, we can view $\rho$ as the background dilaton on the background $\cM^{1,3}_{\tJ}$ in consideration. The effective metric governing the local physics is therefore 
\begin{align}
    G_{\mu\nu} = \sinh(\tau) \eta_{\mu\nu} =\rho^{2/3}\eta_{\mu\nu}\,.
    \ 
\end{align}
This metric
describes a cosmological FLRW space-time \cite{Steinacker:2017vqw} where the dilaton $\rho$ evolves with the cosmic scale parameter $a(t)$ as
\begin{align}
\label{rho-a(t)}
    \rho^2=\sinh^3(\tau)\sim e^{3\tau} \ \propto \ a(t)^2 \ , \qquad a(t) \propto t
\end{align}
in the late-time regime under consideration. This means that although $\rho$ is increasing in time, it can be considered as constant for time scales relevant to the {\em local} scattering processes under consideration here.
We can therefore absorb that conformal factor in the local propagating fields in the kinetic term since corrections of order $\cO(H = \frac{\dot a}{a}$), where $H$ is the Hubble parameter, can be neglected. Then, the local effective metric coincides with the Minkowski metric $\eta_{\mu\nu}$ in the Cartesian coordinates $y^\mu$.
 In particular, we have
\begin{align}\label{tderivative}
    \{\ttb_{\dot\mu},f(y)\}=\sinh(\tau)\p_{\dot\mu}f(y)\,,\qquad \p_{\mu}:=\p/\p y^{\mu}\,,\qquad \p_{\mu}y^{\nu}=\delta_{\mu}{}^{\nu}\,.
\end{align}
Thus, $\ttb_{\dot\mu}$  defines the `momenta' on $\cM^{1,3}$. Further relations between $\ttb$'s and $y$'s are collected in Appendix \ref{app:A}.

\paragraph{Higher-spin modes on $\cM^{1,3}_{\tJ}$.} All higher-spin modes on $\cM^{1,3}_{\tJ}$ are encoded by $S^2_{\tJ}$ and captured by the following expansion
\begin{align}\label{hs-valued-fields}
    \phi^{\ib}(y|u)=\sum_{s}\varphi^{\ib}{}_{\nu(s)}(y)\,u^{\nu(s)}\,,\quad 
        \sa_{\dot\mu}(y|u)=\sum_s\cA_{\dot\mu|\nu(s)}(y)\,u^{\nu(s)}\,,\qquad u^{\nu(s)}=u^{\nu_1}\ldots u^{\nu_s}
\end{align}
where $u^{\mu}$ are the normalized generators of the space-like internal 2-sphere $S_{\tJ}^2$ defined as
\begin{align}
    u^{\mu}:=\frac{\ell_p}{\cosh(\tau)}\ttb^{\mu}\,,\qquad u^{\mu}u_{\mu}=1\,.
\end{align}
Here, the terminology ``space-like'' 2-sphere is justified by recalling that $\ttb_{\mu}y^{\mu}=0$ where 
$y^\mu y_\mu = - R^2\cosh^2(\tau)$ are Cartesian coordinates on the FLRW space-time, 
see Appendix \ref{app:A}. 
Accordingly, the coefficients $\varphi^{\ib}_{\mu(s)}$ and $\cA_{\dot\mu|\nu(s)}$ obey  the following constraints
\begin{subequations}
    \begin{align}
        \varphi^{\ib\mu}{}_{\mu\nu(s-2)}&=0\,,\qquad &y^{\mu}\varphi^{\ib}{}_{\mu\nu(s-1)}&=0\,,\\
        \cA_{\dot\mu|\nu(s-2)\beta}{}^{\beta}&=0\,,\qquad &y^{\beta}\cA_{\dot\mu|\nu(s-1)\beta}&=0\,.
    \end{align}
\end{subequations}
The second constraints on each line above are referred to as the space-like constraints, while the first are simply the trace constraints on the internal indices.

The on-shell dof. of the above higher-spin fields can be counted easily by noticing that 
the space of functions on the internal $S^2_{\tJ}$ can be organized in terms of spherical harmonics, 
so that
$\varphi^{\ib}_{\mu(s)}$ has $6(2s+1)$ physical dof. for each $s\geq 0$.
A similar analysis can be done for $\cA_{\dot\mu|\nu(s)}$ which results in $4(2s+1)$ off-shell dof. To count the physical on-shell dof. of $\cA_{\dot\mu|\nu(s)}$, we shall impose the gauge fixing condition
\begin{align}\label{gaugefix-intertwiner}
\cG(\sa) = \{\ttb_{\dot\mu},\sa^{\dot\mu}\}=0
\end{align}
and quotienting out the linearized gauge transformations $\delta \sa^{\dot\mu}=\{\ttb^{\dot\mu},\Lambda\}$. Then, it can be checked that $\cA_{\dot\mu|\nu(s)}$ carry $2(2s+1)$ on-shell dof. 
Thus, as alluded in the introduction, HS-IKKT is \emph{not} a conventional higher-spin theory.

\paragraph{Local 4-dimensional regime.} 
In this paper, we will consider
the regime where the wavelength of some suitable functions $\varphi(y)$ is much shorter than the cosmic scale denoted as $L_{\rm H}$ but longer than the scale $L_{\rm NC}$, below which the space-time becomes non-commutative. Then the functions $\varphi(y)$ can be effectively treated as $4d$ fields on a local patch of $\cM^{1,3}$.

To characterize this regime more explicitly, let us consider the vector field $\{y^\mu,.\}$ acting on the
spin-1 mode $\phi^{(1)} = \varphi_{\mu}(y) u^{\mu}$: 
\begin{align}
\label{covar-vector-bracket}
    \{y^\mu,\phi^{(1)}\} &= \{y^\mu,\varphi_\nu \}u^\nu + \varphi_\nu\{y^\mu, u^\nu\}\nn\\
    &\approx \theta^{\mu\rho} (\del_\rho\varphi_\nu) u^{\nu}  - \ell_p\frac{\sinh(\tau)}{\cosh(\tau)}\varphi^\mu-\varphi_{\nu}u^{\nu}\theta^{\mu\sigma}\p_{\sigma}\frac{R}{\sqrt{-y^2}}\,
\end{align}
recalling that $y^2=-R^2\cosh^2(\tau)$. Averaging over $S^2_{\tJ}$ in the late-time regime using \eqref{u-combinatoric} 
and \eqref{flat-theta}, we obtain\footnote{The projection $\big|_0$ will be discussed in the next section.}
\begin{align}
  \{y^\mu,\phi^{(1)}\}\big|_0  &\approx  \frac{\ell_p}{3} \Big[y^{\mu}\p^{\rho}\varphi_{\rho}-\cY \varphi^{\mu}\Big] - \ell_p\varphi^{\mu}\,,\qquad \cY:=y^{\sigma}\p_{\sigma}=-y^4\p_4 \ .
\end{align}
where we have approximated $\cosh(\tau)\sim \sinh(\tau)$ at late time. Note that after projecting out the $u$ coordinates, we observe that the last term in \eqref{covar-vector-bracket} is suppressed by a factor of $\sinh^{-1}(\tau)$ and therefore can be neglected. Therefore, the first term dominates for all wavelengths $\l$ shorter than the
curvature scale $L_{\rm H}$ 
 of $H^4$  and  longer than the scale of non-commutativity $L_{\rm NC}$:
\begin{align}\label{asymptotic-scale}
 L_{\rm NC} \ll \l \ll   L_{\rm H} = R\cosh(\tau) 
\end{align}
where \cite{Steinacker:2023myp}
\begin{align}\label{L-cosm-H}
L_{\rm NC}  = \sqrt{L_{\rm H}\, \ell_p} \,,\qquad \ell_p :=\frac{2R}{\tJ} 
\end{align}
measured in Cartesian embedding coordinates $y^{\mu}$.
Here, $\ell_p$ characterizes the internal $S^2_{\tJ}$ via \eqref{S2sphereM13},
while $R$ is the radius of the underlying $H^4$.

The regime \eqref{asymptotic-scale} will be referred to as the {\em local 4-dimensional regime}, where the 2nd term in \eqref{covar-vector-bracket} can be neglected, and the following approximations hold
\begin{subequations}\label{eq:Poisson-oblivion}
    \begin{align}
 \{y^\nu,\phi^{(s)}\} &\approx \{y^\nu,\varphi_{\mu_1 \ldots \mu_s}(y)\}u^{\mu_1} \ldots u^{\mu_s}  \nn\,,\\
 \{\ttb^{\dot\nu},\phi^{(s)}\} &\approx \{\ttb^{\dot\nu},\varphi_{\mu_1 \ldots \mu_s}(y)\}u^{\mu_1} \ldots u^{\mu_s} \, .
 \label{Poisson-asymptot-4D}
\end{align}
\end{subequations}
We will  always consider this regime
for the study of tree-level scattering in the below section.
This means that
$\phi^{(s)} := \varphi_{\mu_1 \ldots \mu_s}(y) u^{\mu_1} \ldots u^{\mu_s}$ can be viewed
effectively as a 4-dimensional $\hs$-valued field on $\cM^{1,3}$,
and the Poisson brackets are oblivious to the $\hs$ generators. We then have the following approximation:
\begin{align}
\{f,g\} &\approx \theta^{\mu\nu} \del_\mu f \del_\nu g \, .
 \label{poisson-approx-x}
\end{align}
Moreover, one can easily verify that the above tensor fields are essentially divergence-free,
\begin{align}
   \del_\mu \{y^\mu,\phi^{(s)}\} &\approx 0 \, .
\end{align}
This follows from the fact that \cite{Steinacker:2022yhs}
\begin{align}
   \label{div-free-2}
   \del_\mu(\rho_M \{y^\mu,\phi^{(s)}\}) = 0 \,,\qquad \rho_M:=\frac{1}{\sinh(\tau)}\,,
\end{align}
which in the late-time regime reduces to 
\begin{align}
    \p_{\mu}\{y^{\mu},\phi^{(s)}\}=0 \ .
\end{align}
In particular, the above consideration
illustrates how the underlying $\hs$ degrees of freedom can be realized in different guises: either in a space-like unitary form \eqref{hs-valued-fields} or in a more conventional form \eqref{covar-vector-bracket} in terms of divergence-free tensor fields. These realizations are related by some field redefinition, which will be discussed in more detail in Subsection \ref{sec:covariant-formulation}.

\section{Vertices, soft limit and scattering in HS-IKKT}\label{sec:3}

In this section, we study 
the soft limit of scattering in the HS-IKKT model. 
We first focus on the space-like formulation of HS-IKKT on $\cM^{1,3}$. In the end of the section, we will also briefly discuss the covariant formulation of HS-IKKT, and show how the standard gravitational coupling between a graviton and two scalar fields arises in this setting. 

\underline{\emph{Remark.}}
In the standard Lorentz-invariant setting,
the sub-leading and sub-sub-leading soft factors for gauge theory and gravity were derived in $4d$ \cite{Casali:2014xpa,Cachazo:2014fwa}. It was later shown that these soft factors can be generalized to any dimension, see e.g. \cite{Bern:2014vva}. Moreover, an infinite set of soft factors can be derived using the Ward-Takahashi identity of large gauge transformation \cite{Hamada:2018vrw}. A generalization of these results to Fronsdal type higher-spin case can be found in \cite{Campoleoni:2017mbt}. See also a recent study of soft theorems in the BFSS matrix model \cite{Miller:2022fvc}.

\subsection{Space-like or unitary formulation}\label{sec:spacelike-formulation}


\paragraph{Kinetic actions.} Let us first consider the kinetic action of the $\hs$-valued fields $\phi^{\ib}$
\begin{align}
    S^{\phi}_2= -\int \mho \,\{\ttb_{\dot\mu},\phi^{\ib}\}\{\ttb^{\dot\mu},\phi_{\ib}\}
    =- \int \mho\, \phi^{\ib}\Box_{1,3}\phi_{\ib}\,,\qquad 
    \Box_{1,3}= -\{t^{\dot\mu},\{t_{\dot\mu},.\}\}
\end{align}
where we have integrated by part.
Similarly, the gauge-fixed kinetic action for the physical $\hs$-valued gauge fields $\sa_{\dot\mu}$, cf. \cite{Steinacker:2023zrb}, reads
\begin{align}\label{eq:gauge-fixed-S2-a}
    S_2^{\sa}=-\int \mho \,
  \sa_{\dot\mu}\Box_{1,3}\sa^{\dot\mu}\, 
\end{align}
in Feynman gauge,
after gauge fixing with 
$\cG(\sa)=0$ cf. \eqref{gaugefix-intertwiner}
and dropping\footnote{Strictly speaking, the kinetic term contains an extra first-derivative term $\{\theta^{\mu\nu},\cA_\nu\}$, which is negligible in the local 4-dimensional regime under consideration. For a careful treatment including this term see \cite{Steinacker:2019awe,Sperling:2019xar}.} the Fadeev-Popov ghosts, since we work at tree level.

Now consider the Box operator $\Box_{1,3}$ acting on the $\hs$-valued fields \eqref{hs-valued-fields}.
As shown in Appendix \ref{app:IRmass},
the Box operator then reduces to the following space-time kinetic operator
\begin{align}
 \Box_{1,3}\phi_{\ib}\, 
 = \Box_{1,3}(\varphi^{\ib}_{\nu(s)}(y) u^{\nu(s)})\, 
 = u^{\nu(s)}\rho^2\Big( \Box_G + \frac{m_s^2}{\rho^2}\Big)\varphi^{\ib}_{\nu(s)}(y)
\end{align}
and similarly for the tangential fluctuations, where\footnote{The specific form of $m^2_s$ with the negative sign should be taken with a pinch of salt, since it may easily change for slightly modified backgrounds.}
\begin{align}
\label{eq:IRmass}
 \Box_G=-\frac{1}{\sqrt{|G|}}\p_{\mu}\Big(\sqrt{|G|}G^{\mu\nu}\Big)\p_{\nu}\,,\qquad m_s^2=-\frac{s}{R^2}\,.  
\end{align}
Observe that the mass terms $\frac{m_s^2}{\rho^2}$ whose scale is set by the cosmic curvature $\rho^{-2} \sim a(t)^{-2}$ (cf. \eqref{rho-a(t)}) are consistent with the $2s+1$ physical degrees of freedom of massive fields. This mass, however, is negligible for the scatterings considerations in the local $4d$ regime. As such, we can drop $m_s^2$ in the following computations,
and replace $\Box_G \to -\eta^{\mu\nu}\del_\mu\del_\nu$ in local Minkowski coordinates.

To express the action in terms of the component fields, 
we can now average over fiber coordinates $\ttb$'s using the projection $[\cdot]_0$ to the space of trivial harmonics\footnote{This is the space where all fiber coordinates are removed/projected out.} on $S^2_{\tJ}$:
\begin{align}\label{u-combinatoric}     [u^{\mu_1}\ldots u^{\mu_{2s}}]_0=\alpha_{2s}\sum [u^{\mu_i}u^{\mu_j}]_0\ldots[u^{\mu_k} u^{\mu_l}]_0,\qquad \alpha_{2s}=\frac{2^s\times s!}{ (2s+1)!}\,,\quad s\geq 1\,
\end{align}
where
\begin{align}\label{u-project}
    [u^{\mu}u^{\nu}]_0=\frac{1}{3}\Big[\eta^{\mu\nu}+\frac{x^{\mu}x^{\nu}}{R^2\cosh^2(\tau)}\Big]=:\hat\kappa^{\mu\nu}\,,\qquad \hat{\kappa}^{\mu\nu}y_{\mu}=0\,.
\end{align}
With this consideration, the kinetic action $S^{\phi}_2$ reads 
\begin{align}\label{kinetic-action-phi1}
    S_2^{\phi}\approx-\int d^4y \sinh(\tau)\,\varphi^{\ib}\Box\varphi_{\ib}-\sum_{s\geq 1}\frac{2\alpha_{2s}}{2s-1}\int d^4y \sinh(\tau) 
    \varphi^{\ib}_{\mu(s)}\Box\varphi^{\ib\mu(s)}\,.
\end{align}
where $\Box = -\del^\mu\del_\mu$ is the standard flat d'Alembertian on Minkowski space. 
In addition, we shall rescale all fields with a factor of $1/\sqrt{\sinh(\tau)}$. Then the fields are canonically normalized (up to a $\cO(1)$ constant), with the kinetic term given by e.g.
\begin{align}\label{kinetic-action-ph2}
    S_2^{\phi}\approx-\int d^4y \,\varphi^{\ib}\Box\varphi_{\ib}-\sum_{s\geq 1}\frac{2\alpha_{2s}}{(2s-1)}\int d^4y \,\varphi^{\ib}_{\mu(s)}\Box\varphi_{\ib}^{\mu(s)}\,.
\end{align}
Repeating the treatment above, the gauge-fixed spacetime kinetic action for higher-spin gauge fields $\cA_{\dot\mu|\nu(s)}$ reads
\begin{align}\label{kinetic-action-A}
    S_2^{\phi}\approx-\int d^4y \,\cA_{\dot\mu}\Box\cA_{\dot\mu}-\sum_{s\geq 1}\frac{2\alpha_{2s}}{2s-1}\int d^4y
    \cA_{\dot\mu|\nu(s)}\Box\cA^{\dot\mu|\nu(s)}\,
\end{align}
in Feynman gauge\footnote{Note that this is compatible with the unitary ``gauge" of the internal modes.}. 

From the above action, we can obtain the corresponding propagators. These propagators will allow computing e.g. the $t$-channel of the simplest scattering amplitude $\cM_4(0,1,1,0)$  between two scalar fields and 2 gauge fields in Subsection \ref{sec:amplitudes}.

To write the above more covariantly 
(cf. section \ref{sec:covariant-formulation}), the frame index of the 
tangential fluctuations could be converted to a covariant index via 
\begin{align}
\label{eq:convert-example}
   \cA_{\dot\mu|\nu} =  E_{\dot\mu}{}^{\sigma}\cA_{\sigma|\nu} \ 
\end{align}
where $\p E^{\dot\mu\mu}$ can be neglected
 in the 
absence of strong gravity. However,
since we are mainly interested in local scattering here, we will often keep the frame index explicit.

\paragraph{Propagators.} Up to normalization, the propagator for $\varphi^{\ib}_{\mu(s)}$ in momentum space and in the unitary gauge reads
\begin{align}\label{propagator-phi-phi}
    \langle \varphi^{\ib}{}_{\mu(s)}(p)\varphi^{\jb}{}_{\nu(s')}(q)\rangle \sim \delta^4(p+q)\delta_{s,s'}\frac{\delta^{\ib\jb}\hatkappa_{\mu\nu}\ldots\hatkappa_{\mu\nu}}{p^2} \,,
\end{align}
where the same label of indices implies symmetrization. Note that all tensors $\hatkappa_{\mu\nu}$ are space-like, and invariant under the space-like $SO(1,3)$ isometry cf. \eqref{u-project}.
Similarly, we obtain the propagator between the spin-$(s+1)$ fields $\cA_{\dot\mu|\nu(s)}$ 
in unitary or space-like gauge 
as 
\begin{align}\label{propagator-spin-2}
    \langle \cA_{\dot\mu|\nu(s)}(p)\cA_{\dot\rho|\sigma(s)}(q)\rangle \sim \delta^4(p+q)\eta_{\dot\mu\dot\rho}\frac{\hatkappa_{\nu\sigma}\ldots\hatkappa_{\nu\sigma}}{p^2}\,.
\end{align}
The above propagators of the gauge fields will allow us to compute the $t$-channel of the simplest 4-point amplitude $\cM_4(0,1,1,0)$ explicitly in Subsection \ref{sec:amplitudes}. Note that we still work with the background coordinates $\ttb^{\dot\mu}$ whenever dealing with the Poisson brackets.

\paragraph{Cubic couplings between $\hs$ scalars and a $\hs$ gauge field.} The relevant cubic vertices that we use to study
scattering in our setting arise from 
\begin{align}\label{main-cubic}
    \cV_3&=\int \mho\,\{\ttb^{\dot\mu},\phi^{\ib}\}\{\sa_{\dot\mu},\phi_{\ib}\}=\int \mho\,\{\phi_{\ib},\{\ttb^{\dot\mu},\phi^{\ib}\}\}\sa_{\dot\mu}\nn\\
    &=\int \mho\,\sinh(\tau)\, \theta^{\rho\sigma}(\p_{\rho}\phi_{\ib})(\p_{\sigma}\p^{\dot\mu}\phi^{\ib})\sa_{\dot\mu}\nn\\
    &\approx\int \mho\sinh(\tau)\theta^{\rho\sigma}(\p_{\rho}\phi_{\ib})(\p_{\sigma}\p^{\mu}\phi^{\ib})\sa_{\mu}
\end{align}
using \eqref{tderivative}. Here, we have assumed gravity is weak so that local Lorentz indices can be converted to spacetime indices if it is needed, cf. \eqref{eq:convert-example}. Due to the presence of $\theta^{\rho\sigma}$, the averaging over fiber coordinates only survives if we have an odd number of $u$'s from the fields. In particular, we get
\begin{subequations}\label{theta-u-average}
    \begin{align}
    [\theta^{\mu\nu}]_0&=0\,,\\
    [\theta^{\mu\nu}u^{\rho}]_0&=\frac{\ell_p\sinh(\tau)}{3\cosh(\tau)}\Big([\eta^{\rho\nu}y^{\mu}-\eta^{\rho\mu}y^{\nu}]+\frac{y_{\beta}\epsilon^{\beta \rho\mu\nu}}{\sinh(\tau)}\Big)\,,\\
     [\theta^{\alpha\beta}u^{\mu_1}\ldots u^{\mu_{2s+1}}]_0&=\alpha_{2s}\sum_{i=1}^{2s+1}[\theta^{\alpha\beta}u^{\mu_i}]_0[uu]_0\ldots[uu]_0\,,\\
     [\theta^{\alpha\beta}u^{\mu_1}\ldots u^{\mu_{2s}}]_0&=0\,.
\end{align}
\end{subequations}
It is useful recalling that
\begin{align}
    \theta^{\mu\nu}&=\frac{\ell_p^2}{\cosh^2(\tau)}\Big(\sinh(\tau)(y^{\mu}\ttb^{\nu}-y^{\nu}\ttb^{\mu})+\eps^{\mu\nu\sigma\rho}y_{\sigma}\ttb_{\rho}\Big)\nn\\
    &=\frac{\ell_p}{\cosh(\tau)}\Big(\sinh(\tau)(y^{\mu}u^{\nu}-y^{\nu}u^{\mu})+\epsilon^{\mu\nu\rho\sigma}y_{\rho}u_{\sigma}\Big)\,
\end{align}
cf. \eqref{mgenerator}. 
In the following, we will keep the gauge potential $\sa_{\dot\mu}$ in the third position in the cubic vertices, and write the vertices as
\begin{align}
    \cV_3^{[\Lambda]}:=\sum_{s_i,s_j,s_k}\cV_3^{(s_i,s_j,s_k)}    \quad \text{such that} \quad s_i+s_j+s_k=\Lambda\,,\quad s_k\geq 1 \ .
\end{align}
Here $s_i,s_j$ denotes the spin of the external fields $\varphi^{\ib}{}_{\mu(s_i)}$ and $\varphi^{\ib}{}_{\mu(s_j)}$, and the third entry $s_k$ is the sum of the internal spin (coming from $\ttb$'s) and the external spin 1 of the $\hs$-valued gauge field $\sa_{\dot\mu}$. We will call $\Lambda$ the total spin. With these notations, we have, for instance:

$\bullet$ \underline{At $\Lambda=1$}, we see that $\cV_3^{[1]}=\cV_3^{(0,0,1)}=0$ since the classical Poisson tensor or $B$ field $[\theta^{\mu\nu}]_0 = 0$ vanishes. 
This is an important result: the interactions vanish identically for all classical fields at cubic order, i.e. without any explicit $\hs$ modes. 
This implies that Lorentz-violating effect of the classical fields can only arise due to the exchange of $\hs$ modes, which is highly suppressed, as we will see.

$\bullet$ \underline{At $\Lambda=2$}, there are three contributions
\begin{align}
    \cV_3^{[2]}=\cV_3^{(1,0,1)}+\cV_3^{(0,1,1)}+\cV_3^{(0,0,2)}
\end{align}
given by 
\small
\begin{subequations}
    \begin{align}
        \cV_3^{(1,0,1)}&=\Xi\int d^4y \Bigg(\Big[(\cY \varphi_{\ib \alpha})(\p^{\alpha}\p^{\mu}\phi^{\ib})-(\p^{\alpha}\varphi_{\ib\alpha})(\cY \p^{\mu}\phi^{\ib})\Big]\cA_{\mu}+\frac{y_{\beta}\epsilon^{\beta\alpha\rho\sigma}}{\sinh(\tau)}(\p_{\rho}\varphi_{\ib\alpha})(\p_{\sigma}\p^{\mu}\phi^{\ib})\cA_{\mu}\Bigg)\,,\\
        \cV_3^{(0,1,1)}&=\Xi \int d^4y \Bigg(\Big[(\cY\phi_{\ib})(\p^{\alpha}\p^{\mu}\varphi^{\ib}{}_{\alpha})-(\p^{\alpha}\phi_{\ib})(\cY \p^{\mu}\varphi^{\ib}{}_{\alpha})\Big]\cA_{\mu}+\frac{y_{\beta}\epsilon^{\beta\alpha\rho\sigma}}{\sinh(\tau)}(\p_{\rho}\phi_{\ib})(\p_{\sigma}\p^{\mu}\varphi^{\ib}_{\alpha})\cA_{\mu}\Bigg)\,, \\
        \cV_3^{(0,0,2)}&=\Xi\int d^4y\Bigg(\Big[(\cY \phi_{\ib})(\p^{\alpha}\p^{\mu}\phi^{\ib})-(\p^{\alpha}\phi_{\ib})(\cY \p^{\mu}\phi^{\ib})\Big]\cA_{\mu\alpha}+\frac{y_{\beta}\epsilon^{\beta\alpha\rho\sigma}}{\sinh(\tau)}(\p_{\rho}\phi_{\ib})(\p_{\sigma}\p^{\mu}\phi^{\ib})\cA_{\mu\alpha}\Bigg)\,
         \label{3-vertices-002}
    \end{align}
    \end{subequations}
\normalsize
assuming canonical normalization for all fields, with\footnote{The relation with the standard coupling of a graviton to $\phi^{\ib}$ will be discussed in subsection \ref{sec:covariant-formulation}.}
\begin{align}
    \Xi:=\frac{\ell_p}{3\cosh(\tau)\sqrt{\sinh(\tau)}}\,,\qquad \cY:=y^{\sigma}\p_{\sigma}\,,\qquad \sigma=0,1,2,3\,.
\end{align}
Note that all these scaling factors can be considered to be locally constant. The presence of the Levi-Civita tensors $\epsilon^{\alpha\beta\gamma\delta}$ in the sub-vertices of $\cV^{[2]}_3$ indicates that HS-IKKT is chiral. Nevertheless, the action is real, with some apparent mild Lorentz violation effect due to the time-like vector field $\cY$. 

This can be  simplified by noticing that in the late time limit, one can effectively write
\begin{align}
 \cY\sim y^0\p_0=R\cosh(\tau)\p_0  \ 
\end{align}
near some reference point $\tp$ with coordinates $y^{\mu}\big|_{\tp}=(y^0,0,0,0)$.
Hence, $\cY$ is essentially the kinetic energy times the factor $R\cosh(\tau)$, which may be treated locally as a large constant.
 Taking into account the scaling factor $\Xi$, we see that
 $\cV_3^{(s_1,s_2,s_3)}$ are suppressed with a factor of $\ell_p\, R/\sqrt{\sinh(\tau)}$. More explicitly,
\small
\begin{subequations}\label{eq:V3L2}
     \begin{align}
        \cV_3^{(1,0,1)}&\approx\frac{\ell_p\,R}{3\sqrt{\sinh(\tau)}}\int d^4y \Bigg(\Big[(\p_0 \varphi_{\ib \alpha})(\p^{\alpha}\p^{\mu}\phi^{\ib})-(\p^{\alpha}\varphi_{\ib\alpha})(\p_0 \p^{\mu}\phi^{\ib})\Big]\cA_{\mu}+\frac{y_{\beta}\epsilon^{\beta\alpha\rho\sigma}}{\sinh(\tau)}(\p_{\rho}\varphi_{\ib\alpha})(\p_{\sigma}\p^{\mu}\phi^{\ib})\cA_{\mu}\Bigg)\,,\\
        \cV_3^{(0,1,1)}&\approx \frac{\ell_p\,R}{3\sqrt{\sinh(\tau)}} \int d^4y \Bigg(\Big[(\p_0\phi_{\ib})(\p^{\alpha}\p^{\mu}\varphi^{\ib}{}_{\alpha})-(\p^{\alpha}\phi_{\ib})(\p_0 \p^{\mu}\varphi^{\ib}{}_{\alpha})\Big]\cA_{\mu}+\frac{y_{\beta}\epsilon^{\beta\alpha\rho\sigma}}{\sinh(\tau)}(\p_{\rho}\phi_{\ib})(\p_{\sigma}\p^{\mu}\varphi^{\ib}_{\alpha})\cA_{\mu}\Bigg)\,, \label{this-one}\\
        \cV_3^{(0,0,2)}&\approx \frac{\ell_p\,R}{3\sqrt{\sinh(\tau)}}\int d^4y\Bigg(\Big[(\p_0 \phi_{\ib})(\p^{\alpha}\p^{\mu}\phi^{\ib})-(\p^{\alpha}\phi_{\ib})(\p_0 \p^{\mu}\phi^{\ib})\Big]\cA_{\mu\alpha}+\frac{y_{\beta}\epsilon^{\beta\alpha\rho\sigma}}{\sinh(\tau)}(\p_{\rho}\phi_{\ib})(\p_{\sigma}\p^{\mu}\phi^{\ib})\cA_{\mu\alpha}\Bigg)\,\label{soft-test}
    \end{align}
\end{subequations}
\normalsize
at late times $\tau \gg 1$. 
These vertices explicitly 
 break Lorentz invariance via the cosmic time-like vector field $\del_0$, which measures the energy of local perturbations in the FLRW frame. However, this breaking is suppressed at late times and low energy.
 Moreover, we will see in section \ref{sec:covariant-formulation} that Lorentz invariance is recovered in the covariant formulation of the model, at least for some of these vertices.
 To understand the associated scale, we note that
 these vertices scale as 
\begin{align}
    \cV_3 \sim \frac{L_{\rm NC}^2}{\cosh(\tau)\sqrt{\sinh(\tau)}}\approx L^2_{\rm NC}\,e^{-\frac{3\tau}{2}} \qquad\text{where}\qquad  L_{\rm NC}  = \sqrt{\ell_p\,R\cosh(\tau)}\,
\end{align}
is the non-commutative length scale
(cf. \eqref{L-cosm-H}).
Thus, the cubic vertices are strongly suppressed in the late-time $\tau\rightarrow \infty$ limit for energies less than 
$\L_{\rm NC} = 1/L_{\rm NC}$.
The relevance of this scale is not surprising, since the interactions arise solely from commutators of the fields.
Beyond this scale, the present semi-classical formalism is no longer applicable and must be replaced by the fully non-commutative and non-local matrix model framework.

To summarize, we have seen that 
the  (higher-spin) cubic vertices \eqref{main-cubic} are proportional to $L^2_{\rm NC}\,e^{-\frac{3\tau}{2}}$, i.e.
\begin{align}
    S\approx S_2^{\phi}+L^2_{\rm NC}\,e^{-\frac{3\tau}{2}}S_3(\phi^{\ib},\phi_{\ib},\sa) \ .
\end{align}
The explicit  $L^2_{\rm NC}$ indicates that
the extra derivatives in $S_3$ are suppressed  by the UV scale $L^2_{\rm NC}$ (note that $S_3$ has dimension [$mass^2$]). Here, the factor $L^2_{\rm NC}\,e^{-\frac{3\tau}{2}}$ is a time-dependent coupling in the local physical regime, which becomes small at late times. 
It is hence interesting to study the soft-limit of scattering amplitudes in 
Lorentzian HS-IKKT, 
which will be the subject of the next subsection.

\paragraph{Cubic couplings between $\hs$-valued gauge fields.} For our purpose of computing the simplest 4-point amplitude $\cM_4(0,1,1,0)$, we will need the cubic vertex between three $\hs$-valued gauge fields $\sa_{\dot\mu}$:
\begin{align}
\label{cubic-YM-vertex-U}
    \cU_3=\int \mho \,\{\ttb^{\dot\mu},\sa^{\dot\nu}\}\{\sa_{\dot\mu},\sa_{\dot\nu}\}\,.
\end{align}
Proceeding as before, the cubic vertices $\{\cU_3^{(1,1,2)},\cU_3^{(1,2,1)},\cU_3^{(2,1,1)}\}$ in the flat late time regime read (note that we suppress contributions associated with the Levi-Civita symbols and convert all frame indices to spacetime ones as before) as
\begin{subequations}\label{eq:U3}
    \begin{align}
        \cU_3^{(1,1,2)}&\approx \frac{1}{3}L_{\rm NC}^2e^{-\frac{3}{2}\tau}\int (\p^{\mu}\cA^{\nu})(\p_0 \cA_{\mu})(\p^{\zeta}\cA_{\nu\zeta})-(\p^{\mu}\cA^{\nu})(\p^{\zeta}\cA_{\mu})(\p_0 \cA_{\nu\zeta})\,,\\
        \cU_3^{(1,2,1)}&\approx \frac{1}{3}L_{\rm NC}^2e^{-\frac{3}{2}\tau}\int (\p^{\mu}\cA^{\nu})(\p_0 \cA_{\mu\zeta})(\p^{\zeta}\cA_{\nu})-(\p^{\mu}\cA^{\nu})(\p^{\zeta}\cA_{\mu\zeta})(\p_0 \cA_{\nu})\,, \\
        \cU_3^{(2,1,1)}&\approx \frac{1}{3}L_{\rm NC}^2e^{-\frac{3}{2}\tau}\int (\p^{\mu}\cA^{\nu\sigma})(\p_0 \cA_{\mu})(\p_{\sigma}\cA_{\nu})-(\p^{\mu}\cA^{\nu\zeta})(\p_{\zeta}\cA_{\mu})(\p_0 \cA_{\nu}) \,.
    \end{align}
\end{subequations}
where we have converted dotted (frame) indices to spacetime indices similarly to \eqref{eq:convert-example}. 


\subsection{Soft limit of higher-spin scatterings}

This subsection discusses the soft limit of scatterings in HS-IKKT. In particular, we study higher-spin soft factors of HS-IKKT theory using the cubic vertex \eqref{main-cubic}. We will see that in contrast to the conventional picture,
there are no non-trivial constraints arising from gauge invariance of the HS-IKKT $S$-matrix in the soft limit. 
We also compute some 3-point scattering amplitudes and observe that they can be non-trivial when the external fields are the would-be-massive modes. Nevertheless, these amplitudes are suppressed by a factor $L_{\rm NC}^2\,e^{-\frac{3}{2}\tau}$ as discussed above. In addition, we show that these amplitudes vanish on-shell upon restricting to their massless sector, where fields are divergence-free. 
Finally, we obtain the fusion rules or spin constraints for the cubic vertices in the current setting, and use them to study the simplest 4-point amplitude between two spin-0 and two gauge fields.

\paragraph{On Weinberg's soft theorem.} It is well-known that gauge invariance of the $S$-matrix can severely constrain the IR physics. This fact was noticed long ago by various authors \cite{Low:1954kd,Gell-Mann:1954wra,Low:1958sn,kazes1959generalized,Gross:1968in,jackiw1968low,Saito:1969lga,Ferrari:1971at} when they studied the soft factors of certain $S$-matrices. One of the most notable examples is Weinberg's soft theorem \cite{Weinberg:1964ew}, where the leading soft factor of the $S$-matrix alone can rule out the existence of an interacting macroscopic massless higher spin field with spin $s>2$. However, the cubic vertices in most of these studies were parity- and Lorentz-invariant. Once parity invariance is relaxed, one can have higher-spin scatterings, as argued in \cite{Tran:2022amg}. In the following, we will see how  Weinberg's soft theorem is avoided in the present framework through Lorentz-violating vertices in the space-like formulation of HS-IKKT.

\paragraph{Setup.} Consider a scattering process of $n$-point scattering amplitudes where the external legs are the $\hs$-valued scalar fields $\phi^{\ib}$ and the emitting soft particle is a $\hs$-valued gauge field $\sa_{\dot\mu}$ cf. \eqref{hs-valued-fields}.

\begin{figure}[ht!]
    \centering
    (\text{a}) \includegraphics[scale=0.3]{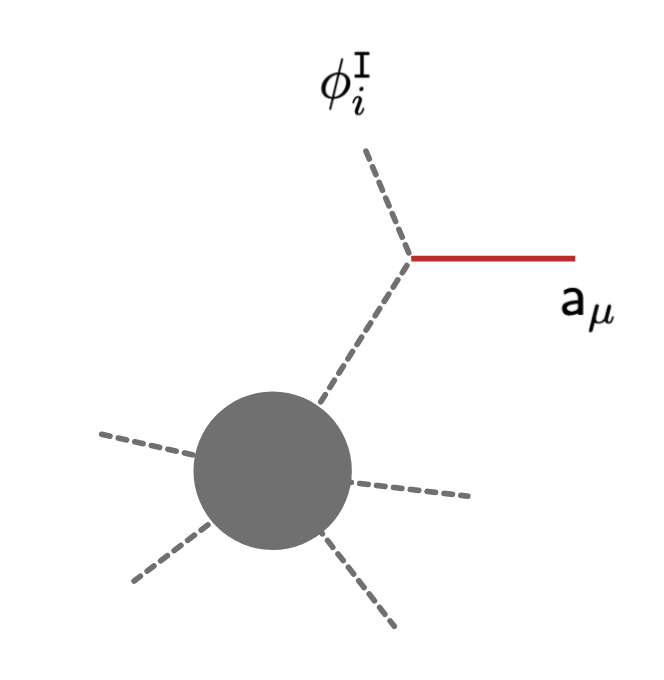}\qquad \text{(b)}
     \includegraphics[scale=0.3]{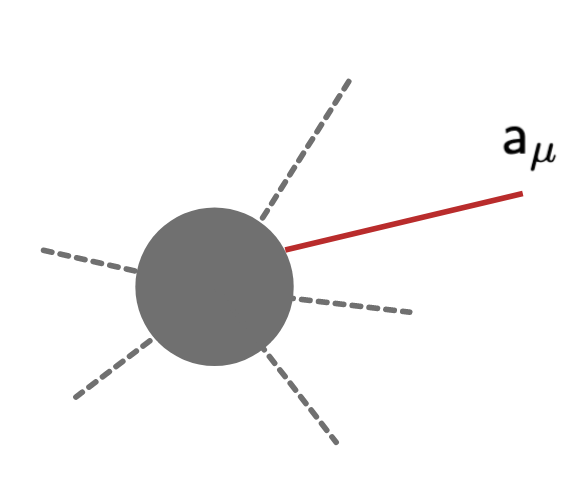}
     \caption{The leading soft factor of the $S$-matrix arise from diagrams where $\sa_{\mu}$ is emitted from an external leg (figure-a above), while sub-leading soft factors arise from diagrams of both types (a) and (b).}
     \label{fig:1}
\end{figure}

Let $\cM_n$ be the $n$-point tree-level scattering amplitude of $n$ $\hs$-valued scalar fields $\phi^{\ib}$ 
whose momenta are $p_i^{\mu}$ [dashed lines]. We will consider $\sa_{\dot\mu}$ as a soft emitting $\hs$-valued gauged field [red] with momenta $k^{\mu}$ from this scattering process (see Figure \ref{fig:1}). Of course, we will really look at each sub-vertex $\cV_3^{(s_1,s_2,s_3)}$ at a fixed total spin $\Lambda$. 

\paragraph{Kinematics.} We denote all external hard momenta of $\phi^{\ib}_i$ as $p_i^{\mu}$, and the soft momenta of $\sa_{\mu}$ as $k_{\mu}$. Here, the indices $i=1,\ldots,n$ are used to label the external particles. We can assume that the polarization tensors $\epsilon^{\ib}_{\mu(s)},\varepsilon_{\dot\mu|\nu(s)}$ are traceless in the space-like indices,
\begin{subequations}\label{kinemtics}
    \begin{align}
   \varphi^{\ib}_{\mu(s)}&: \qquad  &\epsilon^{\ib\nu}{}_{\nu\mu(s-2)}&=0\,,\\
  \cA_{\dot\mu|\nu(s)}&: \qquad & \varepsilon_{\dot\mu|\nu(s-2)\beta}{}^{\beta}&=0\,,\qquad k^{\dot\mu}\varepsilon_{\dot\mu|\nu(s)}=0\,,
\end{align}
\end{subequations}
for all higher-spin modes $(\varphi^{i\ib}_{\mu(s)},\cA_{\dot\mu|\nu(s)})$. 

\paragraph{Gauge transformation.} 
Gauge transformations in the present framework 
 are given by
\begin{align}
    \delta_\xi \sa_{\dot\mu}
    =\sum_s\{\ttb_{\dot\mu},\xi_{\nu(s)}u^{\nu(s)}\}\,,
\end{align}
where in the local $4d$ regime
\begin{align}
    \{\ttb_{\dot\mu},\xi_{\nu(s)}u^{\nu(s)}\}\approx \big(\sinh(\tau)\p_{\dot\mu}\xi_{\nu(s)}+s\tanh(\tau)y_{\dot\mu} \xi_{\nu(s)}\big)u^{\nu(s)}\,.
\end{align}
At late time $\tau$, we can neglect the last term  since $\lim_{\tau\rightarrow \infty}\tanh(\tau)=1$. Thus, to a very good approximation:
\begin{align}
\label{gauge-trafo-explicit}
    \{\ttb_{\dot\mu},\xi_{\nu(s)}\}\approx \sinh(\tau)\p_{\dot\mu}\xi_{\nu(s)}\,.
\end{align}
Hence, the gauge transformation for the $\hs$-valued gauge field $\sa_{\dot\mu}$ in the $4d$ local regime reads
\begin{align}
    \delta_\xi \sa_{\dot\mu}\approx\sum_s\sinh(\tau)\p_{\dot\mu}\xi_{\nu(s)}u^{\nu(s)}\,.
\end{align}
As a result,
\begin{align}\label{pre-soft-V}
    \delta_{\xi} \cV_3=\sum_s\int \mho \,\sinh(\tau)^2\theta^{\rho\sigma}(\p_{\rho}\phi_{\ib})(\p_{\sigma}\p^{\mu}\phi^{\ib})\p_{\mu}\xi_{\nu(s)}u^{\nu(s)}\,.
\end{align}
This allows us to work with $\hs$-valued functions $(\phi^{\ib}(y,u),\xi(y,u))$ when studying gauge invariance of the $S$-matrix in the soft limit.
Upon integrating out $u$'s from \eqref{pre-soft-V}, this boils down to the standard gauge transformation on spacetime
\begin{align}
    \delta_\xi \cA_{\dot\mu|\nu(s)}\approx\p_{\dot\mu}\xi_{\nu(s-1)}
\end{align}
where we recall that the IR masses cf. \eqref{eq:IRmass} associated to $\cA_{\dot\mu|\nu(s)}$ are approximately zero in the local $4d$ regime.

\paragraph{Soft factors in Lorentz-invariant field theory.} In general, the soft factors of $n$-point scattering amplitudes $\cM_n$ are defined by 
\begin{align}
    \cM_{n+1}(p_1,\ldots,p_n;k)=\Big(\cS_{(\ell)}+\cS_{(\ell+1)}+\cS_{(\ell+2)}+\ldots\Big)\cM_n(p_1,\ldots,p_n)\,
\end{align}
where $\ell$ is for leading, and
\begin{align}
    \cS_{(n)}=\sum_{i=1}^n\frac{f_{(n+1)}(k,\varepsilon,\tilde\varepsilon;p_i,J_i,\epsilon_i)}{k\cdot p_i}\,,\qquad J_i^{\mu\nu}:=p_i^{\mu}\frac{\p}{\p p_{i\nu}}-p_i^{\nu}\frac{\p}{\p p_{i\mu}}
\end{align}
is the soft factor of order $n$ in the soft momenta $k^{\mu}$ and $f_{(n+1)}$ are some functions of order $(n+1)$ in the soft momentum $k^{\mu}$. Here, $1/(k\cdot p_i)$ is the propagator of the higher-spin fields $\varphi^{\ib}_{\mu(s)}$ linking $\cM_n$ and the cubic vertices. Note that only the gauge fields need to be massless, while the $\hs$-valued scalar fields can be massive in obtaining the above soft factors.

Imposing gauge invariance on the amplitudes, all soft theorems can be extracted from
\begin{align}
    \sum_{n\geq \ell} k_{\mu}\cS^{\mu\nu(s-1)}_{(n)}=0\,,
\end{align}
where $s$ denotes the \emph{spin} of the emitting particle encoded in the $\hs$-valued gauge field $\sa_{\dot\mu}$. Note that all sub-leading soft factors can be obtained by systematically trading each hard external momenta $p_i^{\mu}$ in $f_{(\ell+1)}(k,\varepsilon,\tilde\varepsilon;p_i,\epsilon_i)$ for a factor of $k_{\nu}J_i^{\nu\mu}$ cf. \cite{Cachazo:2014fwa,Bern:2014vva}. In deriving this result, all vertices in the standard quantum field theory approach are Lorentz-invariant. Moreover, since the angular momentum $J^{\mu\nu}$ is anti-symmetric, there is no need to go beyond $\cS_{(\ell+2)}$ to constrain the $S$-matrix due to the fact that $\cS_{(\ell+2)}$ is trivially gauge invariant due to symmetry. As such, $\cS_{(\ell)}$ and $\cS_{(\ell+1)}$ are usually all we need to determine IR physics. From this discussion, it is clear that once we obtain $\cS_{(\ell)}$, all soft theorems can be determined explicitly.

\paragraph{Soft limit in HS-IKKT scatterings.} We can now return to our setting where the cubic vertex \eqref{main-cubic}  mildly violates Lorentz invariance 
and is not parity invariant. Let us consider a process where there are emission of a soft particle of fixed spin, say $s_3$, from the external legs of an $(n-1)$-point amplitude $\cM_{n-1}$ (see diagram (a) in Fig \ref{fig:1}). We will assume that the propagator, the external field and the soft emitting particle from $\cV_3^{(s_1,s_2,s_3)}$ carry momenta $\{p_i+k,p_i;k\}$, respectively. 

Considering a gauge transformation and plugging in plane-wave solutions to $\delta_{\xi}\cV_3^{(s_1,s_2,s_3)}$, we first observe that all terms come with the Levi-Civita symbols $\epsilon^{\mu\nu\rho\sigma}$ vanish due to symmetry in the soft limit. Furthermore:

$\bullet$ \underline{At $\Lambda=1$}, 
\begin{align}
    \delta_\xi\cV_3^{(0,0,1)}=0\,
\end{align}
since the vertex vanishes identically.
Hence, no constraint arises in this case. 

$\bullet$ \underline{At $\Lambda=2$}, non-trivial vertices are in \eqref{eq:V3L2}. Let us consider \eqref{soft-test} as an example. The gauge transformation of this vertex results in, 
\begin{align}\label{delta-V002}
   \delta_\xi \cV_3^{(0,0,2)}\approx L_{\rm NC}^2e^{-\frac{3}{2}\tau}\sum_i(p_i\cdot k)\Big[(E_i+k)p_i^{\alpha}-(p_i+k)^{\alpha}E_i\Big]\,.
\end{align}
where $E_i:=p_{i0}$ is the \emph{energy} of the $i$th particle 
in the locally flat regime. As usual, the associated soft factor of the above can be obtained by dividing \eqref{delta-V002} with the propagator. Thus,
\begin{align}
    \cS_{(\ell)}[\cV_3^{(0,0,2)}]\approx L_{\rm NC}^2e^{-\frac{3}{2}\tau}\sum_i\Big[(E_i+k)p_i^{\alpha}-(p_i+k)^{\alpha}E_i\Big]\,.
\end{align}
It is easy to see that, $\cS_{(\ell)}[\cV_3^{(0,0,2)}]$ vanishes identically in the soft limit, i.e.
\begin{align}\label{soft-theorem}
   \lim_{k\rightarrow 0}\cS_{(\ell)}[\cV_3^{(0,0,2)}]\approx L_{\rm NC}^2e^{-\frac{3}{2}\tau}\sum_iE_{i} \Big[p_i^{\mu} -p_i^{\mu}\Big]=0\,.
\end{align}
Therefore, no non-trivial constraints arise from the soft factors of the $S$-matrices. This is due to the non-standard, Lorentz-violating, and antisymmetric structures of the vertices. 
We show below that these vertices can in fact lead to non-trivial scatterings. 

The story for cubic vertices with other higher-spin fields when $\Lambda\geq 4$ is completely analogous.\footnote{The readers can find the cubic vertices at $\Lambda=4$ in Appendix \ref{app:B}.} Thus, the soft limit associated to the cubic vertex \eqref{main-cubic} in HS-IKKT simply provides the trivial relation \eqref{soft-theorem}. 


As a final remark, we note that to have a local higher-spin theory with propagating degrees of freedom, one usually needs to abandon either unitarity as in the case of conformal higher-spin gravity \cite{Tseytlin:2002gz,Segal:2002gd,Bekaert:2010ky,Basile:2022nou}, or parity invariance as in the case of (quasi-) chiral higher-spin theories \cite{Metsaev:1991mt,Metsaev:1991nb,Ponomarev:2016lrm,Ponomarev:2017nrr,Metsaev:2018xip,Metsaev:2019dqt,Metsaev:2019aig,Tsulaia:2022csz,Krasnov:2021nsq,Adamo:2022lah}.\footnote{Note that these theories have simple $S$-matrices \cite{Joung:2015eny,Beccaria:2016syk,Roiban:2017iqg,Skvortsov:2018jea,Tran:2022amg}.} Furthermore, it is also well-known that the construction of higher-spin theories using standard Noether couplings cannot surpass no-go theorems (see the discussion in Appendix \ref{app:noether} where we consider a non-commutative version of Noether couplings for higher-spin fields). However, in the present case of Lorentzian HS-IKKT \cite{Steinacker:2019awe}, the price to pay for a \emph{local} higher-spin theory is somewhat different. Due to the presence of the cosmic vector field $\cY$ in the cubic vertices, Lorentz invariance is not manifest, at least in the unitary formulation. This allows us to surpass the usual no-go theorems without sacrificing unitarity.\footnote{There is a class of $3$-dimensional higher-spin gravities \cite{Blencowe:1988gj,Bergshoeff:1989ns,Pope:1989vj,Fradkin:1989xt,Campoleoni:2010zq,Henneaux:2010xg,Gaberdiel:2010pz,Gaberdiel:2012uj,Gaberdiel:2014cha,Grigoriev:2019xmp,Grigoriev:2020lzu} which is local, unitary, and has a truncated spectrum of higher-spin fields. However, these theories do not have propagating degrees of freedom.}


\subsection{Some 3-point and 4-point higher-spin amplitudes}\label{sec:amplitudes}

Since the HS-IKKT model on $\cM^{1,3}$ contains Lorentz-violating interactions for the $\hs$ modes, which include extra degrees of freedom corresponding to the would-be-massive modes, non-trivial 3-point scattering amplitudes do arise for the $\hs$ modes. This is due to the fact that
\begin{align}
    p_i^{\mu}\epsilon_{i\mu\nu(s-1)}\neq 0\,,\qquad k^{\nu}\varepsilon_{\dot\mu|\nu\beta(s-1)}\neq 0\,,
\end{align}
since the fields are the would-be-massive modes, which are a priori \emph{not} divergence-free. If we want to zoom in to the strictly massless limit, then we should impose the following additional constraints on the above polarizations
\begin{align}\label{eq:additional-constraints}
    p_i^{\mu}\epsilon_{i\mu\nu(s-1)}= 0\,,\qquad k^{\nu} \varepsilon_{\dot\mu|\nu\beta(s-1)}= 0\,,\qquad y^{\dot\mu}\varepsilon_{\dot\mu|\nu(s)}=0\,.
\end{align}
In this case, all amplitudes vanish on-shell, in agreement with the standard story of QFT. It can then be checked that all fields now have precisely 2 on-shell physical dof.

Let $\cM_3^{(s_1,s_2,s_3)}=\cM_3(p_1,\epsilon_1,p_2,\epsilon_2;k_3,\varepsilon_3)$ be the 3-point amplitudes between two higher-spin fields $\varphi^{\ib}_{\mu(s)}$ and a gauge field $\cA_{\dot\mu|\nu(s)}$. One can check that 

$\bullet$ \underline{At $\Lambda=1$}, the 3-point amplitude vanishes $\cM_3^{(0,0,1)}=0$. 

$\bullet$ \underline{At $\Lambda=2$}, we have
\small
\begin{subequations}
    \begin{align}
        \cM_3^{(1,0,1)}&\sim \int d^4 y e^{\im y\cdot (p_1+p_2+k_3)}\Big[(y\cdot p_1)(p_2\cdot \epsilon_1)-(p_1\cdot \epsilon_1)(y\cdot p_2)\Big](p_2\cdot\varepsilon_3)-(1\leftrightarrow 2)\,,\\
        \cM_3^{(0,1,1)}&\sim \int d^4 y e^{\im y\cdot (p_1+p_2+k_3)}\Big[(y\cdot p_1)(p_2\cdot \epsilon_2)-(p_1\cdot\epsilon_2)(y\cdot p_2)\Big](p_2\cdot\varepsilon_3)-(1\leftrightarrow 2)\,,\\
        \cM_3^{(0,0,2)}&\sim \int d^4 y e^{\im y\cdot (p_1+p_2+k_3)}\Big[(y\cdot p_1)(p_2^{\alpha}p_2^{\alpha}\varepsilon_{3\,\alpha(2)})-(p_1^{\alpha}p_2^{\alpha}\varepsilon_{3\,\alpha(2)})(y\cdot p_2)\Big]-(1\leftrightarrow 2)\,.
    \end{align}
\end{subequations}
\normalsize
Note that the above expressions make sense in the local $4d$ regime at late times, where plane-wave solutions do exist, leading to the usual  Dirac delta function $\delta^4(p_1+p_2+k_3)$ corresponding to (local) momentum conservation. Moreover, when swapping the external legs, there is a minus sign coming from the original Poisson structure on $\P^{1,2}$ cf. \eqref{main-cubic}. Treating $\cY = y^\mu \del_\mu\sim\cosh(\tau)\p_0$, we get
\begin{subequations}
    \begin{align}
        \cM_3^{(1,0,1)}&\sim \delta^4(p_1+p_2+k_3)\Big[ E_1(p_2\cdot \epsilon_1)-(p_1\cdot \epsilon_1)E_2\Big](p_2\cdot\varepsilon_3)-(1\leftrightarrow 2)\,,\\
        \cM_3^{(0,1,1)}&\sim \delta^4(p_1+p_2+k_3)\Big[E_1(p_2\cdot \epsilon_2)-(p_1\cdot\epsilon_2)E_2\Big](p_2\cdot\varepsilon_3)-(1\leftrightarrow 2)\,,\\
        \cM_3^{(0,0,2)}&\sim \delta^4(p_1+p_2+k_3)\Big[E_1(p_2^{\alpha}p_2^{\alpha}\varepsilon_{3\,\alpha(2)})-(p_1^{\alpha}p_2^{\alpha}\varepsilon_{3\,\alpha(2)})E_2\Big]-(1\leftrightarrow 2)\,.
    \end{align}
\end{subequations}
\normalsize
where $E_i:=p_{i0}$. All these 3-point amplitudes are suppressed by a factor of $L_{\rm NC}^2\,e^{-\frac{3}{2}\tau}$, which is not written explicitly. 

As pointed out above, when external fields are would-be-massive modes, some of the above amplitudes are non-trivial. However, we notice that in the collinear limit where $p_1^{\mu}\propto p_2^{\mu}$, as well as the soft limit, all of the above amplitudes go to zero. This applies, in particular, to the on-shell massless modes. Furthermore, in the massless sector where fields satisfy the additional constraints \eqref{eq:additional-constraints}, then 
the above amplitudes also vanish.

$\bullet$ \underline{At $\Lambda=4$}, we have for instance (explicit forms of all sub-vertices at $\Lambda=4$ can be found in Appendix \ref{app:B})
    \begin{align}
        \cM_3^{(0,0,4)}&\sim 0\,.\\
    \end{align}
etc. We refrain from exhibiting further 3-point scattering amplitudes for higher-spin fields and leave it as an exercise for enthusiastic readers. Note that it would be interesting to see how much the Lorentz-violating vertices considered in this work deviate from the standard Lorentz-invariant cubic vertices in $4d$  for massless/massive fields classified by Metsaev in the light-cone gauge \cite{Metsaev:2022yvb}.

\paragraph{Fusion rules.} Recall that the cubic vertex \eqref{main-cubic} only survives for $\Lambda\in 2\N^+$. The local $SO(3)$ invariance then leads to
the following fusion rules or spin constraints
\begin{align}\label{fusion-rule}
    s_3\leq s_1+s_2+2\,,\qquad 
    s_1\leq s_2+s_3+1\,,\qquad s_2\leq s_1+s_3+1\,
\end{align}
for $s_1+s_2+s_3=\Lambda\in 2\N^+$ and $s_1,s_2\geq 0\,,\ s_3\geq 1$. 
This means that only a small number of exchange modes can occur when the external fields have low spins.

\paragraph{Four-point $\cM_4(0,1,1,0)$ amplitude.} 
Let us apply the above fusion rules to study an explicit example of the 4-point amplitude $\cM_4(0,1,1,0)$. This amplitude consists of $s$-, $t$-, $u$-channels and the contact interaction:
\begin{align}
    \cM_4=\parbox{80pt}{\includegraphics[scale=0.25]{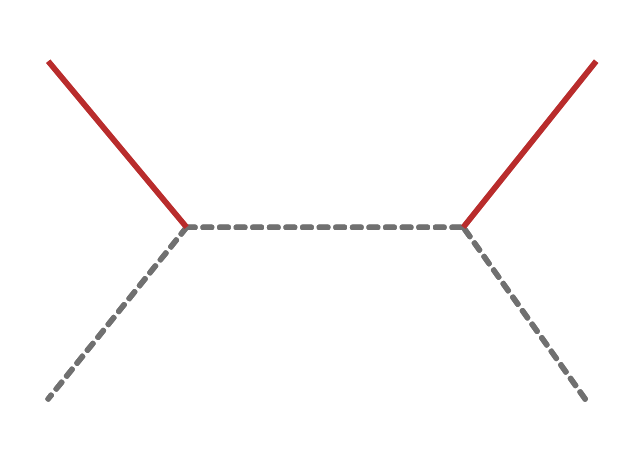}}+\parbox{60pt}{\includegraphics[scale=0.25]{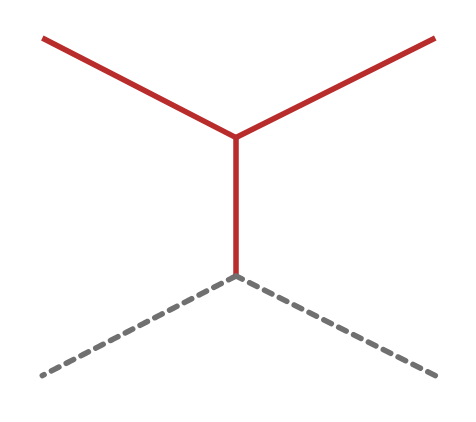}}+\parbox{60pt}{\includegraphics[scale=0.25]{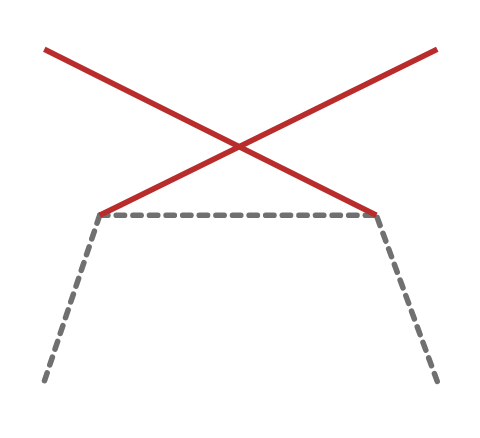}}+\parbox{80pt}{\includegraphics[scale=0.25]{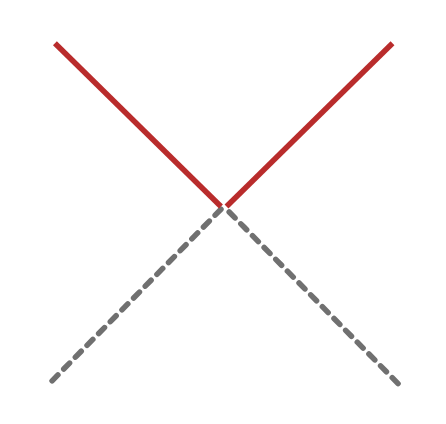}}
\end{align}
where red denotes the gauge fields and dash lines are 
scalar fields. 

On the one hand, using the spin constraints \eqref{fusion-rule}, it follows that the only field that propagates in the exchange of
the $s$- and $u$-channels in $\cM_4(0,1,1,0)$ is $\varphi^{\ib}_{\mu}$. Indeed, we have $\cV^{(s_1\geq 3,0,1)}=0$ and $\cV^{(0,s_2\geq 3,1)}=0$ (see Appendix \ref{app:B}). On the other hand, the $t$-channel receives contributions from the cubic vertices $\cV_3^{(0,0,2)}$, and the set of cubic vertices $\{\cU_3^{(1,1,2)},\cU_3^{(1,2,1)},\cU_3^{(2,1,1)}\}$.\footnote{Notice that when the exchange is a spin-1 gauge field, the $t$-channel vanishes due to the fact that $\cU_3^{(1,1,1)}=0$ and $\cV_3^{(0,0,1)}=0$ using $[\theta^{\mu\nu}]_0=0$.} 

Next, we also need the structure of the (quartic) contact term: 
\begin{align}
\label{quartic-vertex-aapp}
    \cU_4=\frac{1}{2}\int \mho \,\{\sa^{\dot\mu},\phi^{\ib}\}\{\sa_{\dot\mu},\phi_{\ib}\}\approx\frac{1}{2}\int \mho \,\theta^{\rho\sigma}\theta^{\alpha\beta}(\p_{\rho}\sa^{\mu}\p_{\sigma}\phi^{\ib})(\p_{\alpha}\sa_{\mu}\p_{\beta}\phi_{\ib})\,.
\end{align}
Using \eqref{flat-theta}, we can average  
\begin{align}\label{projectionP-munu}
    [\theta^{\mu_1\nu_1}\theta^{\mu_2\nu_2}]_0&\approx \frac{\ell_p^2R^2\sinh^2(\tau)}{3\cosh^2(\tau)}\Big(P^{\mu_1\mu_2}P^{\nu_1\nu_2}-P^{\mu_1\nu_1}P^{\nu_1\mu_2}\Big)\,,\qquad P^{\mu\nu}:=\eta^{\mu\nu}+\frac{y^{\mu}y^{\nu}}{R^2}\,.
\end{align}
This gives 
\small
\begin{align}
    \cU_4^{(0,1,1,0)}\approx\frac{\ell_p^2R^2}{6}e^{-3\tau}&\int \Bigg(\Big[(\p_{\mu}\phi^{\ib})(\p_{\nu}\cA^{\alpha})(\p^{\mu}\phi_{\ib})(\p^{\nu}\cA_{\alpha})-(\p_{\mu}\phi^{\ib})(\p_{\nu}\cA^{\alpha})(\p^{\nu}\phi_{\ib})(\p^{\mu}\cA_{\alpha})\Big] \nn\\
    &+\frac{1}{R^2}\Big[(\cY \phi^{\ib})(\p_{\nu}\cA^{\alpha})(\cY \phi_{\ib})(\p^{\nu}\cA_{\alpha})+(\p_{\mu}\phi^{\ib})(\cY \cA^{\alpha})(\p^{\mu}\phi_{\ib})(\cY \cA_{\alpha})\Big]\nn\\
    &-\frac{1}{R^2}\Big[(\cY \phi^{\ib})(\p_{\nu}\cA^{\alpha})(\p^{\nu}\phi_{\ib})(\cY \cA_{\alpha})+(\p_{\mu}\phi^{\ib})(\cY \cA^{\alpha})(\cY\phi_{\ib})(\p^{\mu}\cA_{\alpha})\Big]\Bigg)\,.
\end{align}
\normalsize
Since terms associated with the time-like vector field $\cY$ are dominant, we can approximate the above to
\small
\begin{align}
    \cU_4^{(0,1,1,0)}\approx\frac{\ell_p^2}{6}e^{-3\tau}\int \Bigg(&+\Big[(\cY \phi^{\ib})(\p_{\nu}\cA^{\alpha})(\cY \phi_{\ib})(\p^{\nu}\cA_{\alpha})+(\p_{\mu}\phi^{\ib})(\cY \cA^{\alpha})(\p^{\mu}\phi_{\ib})(\cY \cA_{\alpha})\Big]\nn\\
    &-\Big[(\cY \phi^{\ib})(\p_{\nu}\cA^{\alpha})(\p^{\nu}\phi_{\ib})(\cY \cA_{\alpha})+(\p_{\mu}\phi^{\ib})(\cY \cA^{\alpha})(\cY\phi_{\ib})(\p^{\mu}\cA_{\alpha})\Big]\Bigg)\,.
\end{align}
\normalsize
where we have normalized $\cU_4$ appropriately. 
Following the same line of argument as above, this quartic vertex is also suppressed in the late time regime. However, note that the quartic vertex is non-trivial even with the lowest possible external spins since $[\theta\theta]_0\neq 0$. 
Furthermore, unlike in Fronsdal-type higher-spin theories (see e.g. \cite{Bekaert:2015tva}), the quartic higher-spin vertices of Lorentzian HS-IKKT are local, at least in the unitary formulation. This remarkable property of the model allows us to use standard QFT techniques to compute the 4-point amplitudes, which are strongly suppressed.

The ingredients to compute the 4-point amplitude $\cM_4(0,1,1,0)$, where external fields are positioned clock-wise, are summarized as follows: 
\begin{itemize}
    \item[(i)] The propagators \eqref{propagator-phi-phi}, \eqref{propagator-spin-2} for $\varphi^{\ib}_{\alpha}$ and $\cA_{\mu|\nu}$, respectively.
    \item[(ii)] The cubic vertices $\{\cV_3^{(0,1,1)}$,$\cV_3^{(1,0,1)}$,$\cV_3^{(0,0,2)}$, $\cU_3^{(1,1,2)}$, $\cU_3^{(1,2,1)}$, $\cU_3^{(2,1,1)}\}$
    . Their explicit forms can be found in Subsection \ref{sec:spacelike-formulation}.
    \item[(iii)] The quartic vertex $\cU_4^{(0,1,1,0)}$.
\end{itemize}
Explicitly, the individual contributions to $\cM_4(0,1,1,0)$ are as follows:

$\bullet$ \underline{$s$-channel:}
\begin{align}
    \cM_s\approx \frac{L^4_{\rm NC}}{9s}e^{-3\tau}\Big(&+2(p_1\cdot \epsilon_2)\big[E_1p_2^{\alpha}-E_2p_1^{\alpha}\big]+(p_2\cdot \epsilon_2)\big[E_1p_2^{\alpha}-E_2p_1^{\alpha}\big]\Big)\nn\\
    \times \hatkappa_{\alpha\beta}\Big(&+2(p_4\cdot \epsilon_3)\big[E_4p_3^{\beta}-E_3p_4^{\beta}\big]+(p_3\cdot \epsilon_3)\big[E_4p_3^{\beta}-E_3p_4^{\beta}\big]\Big)\,.
\end{align}
\normalsize

$\bullet$ \underline{$t$-channel:}
\begin{align}
    \cM_t\approx &+\frac{L_{\rm NC}^4}{9t}e^{-3\tau}\Big(+(p_2\cdot \epsilon_3)\epsilon_2^{\nu}\big[E_2p_3^{\zeta}-E_3p_2^{\zeta}\big]+(\epsilon_2\cdot \epsilon_3)p_2^{\nu}\big[E_3p_2^{\zeta}-E_2p_3^{\zeta}\big]\nn\\
    &\qquad \qquad +\big[(p_2+p_3)\cdot\epsilon_2\big]\epsilon_3^{\nu}\big[E_3p_2^{\zeta}-E_2p_3^{\zeta}\big]-(2\leftrightarrow 3)\Big)\nn\\
    &\qquad \qquad \times \eta_{\beta\nu}\kappa_{\zeta\alpha}\Big(\big[E_1p_4^{\alpha}-E_4p_1^{\alpha}\big]p_4^{\beta}-(1\leftrightarrow 4)\Big)\,.
\end{align}
As a reminder, we have used the frame under the assumption that gravity is sufficiently weak to convert the pair of contracted frame indices to spacetime one.

$\bullet$ \underline{$u$-channel:}
\begin{align}
    \cM_u\approx\frac{L^4_{\rm NC}}{9u\,}e^{-3\tau}\Big(&+2(p_1\cdot \epsilon_3)\big[E_1p_3^{\alpha}-E_3p_1^{\alpha}\big]+(p_3\cdot \epsilon_3)\big[E_1p_3^{\alpha}-E_3p_1^{\alpha}\big]\Big)\nn\\
    \times\hatkappa_{\alpha\beta}\Big(&+2(p_4\cdot \epsilon_2)\big[E_4p_2^{\beta}-E_2p_4^{\beta}\big]+(p_2\cdot \epsilon_2)\big[E_4p_2^{\beta}-E_2p_4^{\beta}\big]\Big)\,.
\end{align}
\normalsize

$\bullet$ \underline{contact term:}
\small
\begin{align}
    \cM_c&\approx \frac{L_{\rm NC}^4}{6}e^{-3\tau}\Bigg[(\epsilon_2\cdot\epsilon_3)\Big((p_2\cdot p_3)(E_1E_4+E_2E_3)-(p_1\cdot p_3)(E_1E_3+E_2E_4)\Big)\pm \binom{1\leftrightarrow 4}{2\leftrightarrow 3}\Bigg]\,. 
\end{align}
\normalsize
Here $(s,t,u)$ are the standard Mandelstam variables, and we recall that all $\hatkappa_{\mu\nu}$ tensors in the above expressions are space-like cf. \eqref{u-project}. In addition, we have suppressed the overall Dirac delta function for momentum conservation, and used $\cY\sim \cosh(\tau)\p_0$ as above. Again, we note that even though the above contributions to the 4-point amplitudes $\cM_4(0,1,1,0)$ in Lorentzian HS-IKKT are non-trivial, they are strongly suppressed in the late time regime.\footnote{It would be interesting to compute higher-point amplitudes for higher-spin fields. However, with the vectorial description used in this paper, the final expressions will be quite involved. For this reason, having a spinorial description for Lorentzian HS-IKKT will be more efficient. We leave this study for future work.}

\subsection{Covariant higher-spin fields and gravitational couplings}
\label{sec:covariant-formulation}

So far, we used the space-like representation \eqref{hs-valued-fields} of fields, which is useful for amplitude calculations but obscures Lorentz invariance.
This subsection briefly discusses the covariant representation of higher-spin fields on $\cM^{1,3}$. 
In particular,
we will consider the gravitational coupling between a covariant graviton and two scalar fields 
given by \eqref{main-cubic}, which looks rather strange in the space-like formulation.
We will see that such vertices can 
be rewritten in a perfectly standard form using the covariant field representations.

\paragraph{Covariant higher-spin fields.} We have seen that the space-like realization of the $\hs$-valued fields interact via Lorentz-violating vertices. 
On the other hand, the same degrees of freedom can also be represented in a more conventional form in terms of ``covariant" rank-$s$ tensor fields, which have the form 
\begin{align}\label{covariant-mapping}
    \Phi^{\ib}_{\mu(s)} = \{y_{\mu_1},\{\ldots,\{y_{\mu_s},\phi^{\ib}_{(s)}\} \ldots \}\}\big|_0
\end{align}
where $\phi^{\ib}_{(s)}:=\varphi^{\ib}_{\nu(s)}u^{\nu(s)}$ arises from transversal matrices. 
These field transform as tensors under volume-preserving diffeos 
\cite{Steinacker:2020xph,Steinacker:2024unq}
\begin{align}
    \delta_\L \Phi^{\ib}_{\mu(s)} = \cL_\xi \Phi^{\ib}_{\mu(s)}, \qquad \xi^\mu = \{\L,y^\mu\}
\end{align}
where $\xi^\mu$ is a divergence-free vector field generated by the gauge parameter $\Lambda$.
Here $\Phi^{\ib}_{\mu(s)}$ are totally symmetric and divergence-free tensor fields, i.e. \cite{Steinacker:2022yhs}
\begin{align}\label{div-free-tensor}
    0=\p_{\mu}(\rho_M \Phi^{\mu\nu(s-1)})\,,\qquad \rho_M=(\sinh(\tau))^{-1}\,
\end{align}
cf. \eqref{div-free-2}, and no longer space-like.
Note that the standard partial derivative $\p_{\mu}$ on $\cM^{1,3}$ is related to the tangential derivative $\eth_a$ on $H^4_{\tJ}$ via.
\begin{subequations}
    \begin{align}
    \p_{\mu}:&=\eth_{\mu}-\frac{y_{\mu}}{y_4}\eth_4\,,\\
    \eth^af(y):&=-\frac{1}{\ell_p^2R^2}\theta^{ab}\{y_b,f(y)\}\,,\qquad a=0,1,\ldots,4\,,\\
    \eth_4:&=\frac{x^{\nu}}{\ell_p^2R^2}\{\theta_{4\nu},.\}=-\frac{\ell_p}{R^2}x_{\mu}\{\ttb^{\mu},.\}=-\frac{\ell_p}{R^3}x_4\cY\,,\qquad \cY:=x^{\mu}\p_{\mu}\,.
\end{align}
\end{subequations} 
Here, $\eth^a$ is a tangential derivative in the sense that $x^a\eth_a=0$. We restrict ourselves to the local physical regime where  
\begin{align}
    \theta^{\mu\nu}
    \approx \ell_p^2(y^{\mu}u^{\nu}-y^{\nu}u^{\mu})\,
\end{align}
using $\cosh(\tau)\sim \sinh(\tau)$ for $\tau\rightarrow \infty$. In this regime, 
 $\{y_{\mu},-\}$ acts trivially on the $u$ generators, so that the covariant higher-spin fields $\Phi^{\ib}_{\mu(s)}$ are obtained using \eqref{eq:PoissonM13} as
\begin{align}
    \Phi^{\ib}_{\mu(s)}\approx \big[\theta_{\mu_1}{}^{\sigma_1}\ldots\theta_{\mu_{s}}{}^{\sigma_{s}}u^{\nu_1}\ldots u^{\nu_{s}}\big]_0\p_{\sigma_1}\ldots\p_{\sigma_{s}}\varphi^{\ib}_{\nu(s)} \ .
\end{align}
It can be checked that $\p^{\mu}\Phi^{\ib}_{\mu\nu(s-1)}=0$. This provides a map from the space-like formulation to the
covariant formulation of $\hs$ fields in Lorentzian HS-IKKT theory. 

To illustrate the relevance of the above covariant tensors, consider again the case of spin 2 fluctuations arising from the first higher-spin mode of $\sa_\mu$, i.e. 
\begin{align}
    \sa_{\dot\mu}^{(1)} = \cA_{\dot\mu|\nu}(y) u^\nu \ .
\end{align}
This determines the following ``covariant" 
graviton as a derived object of the $\hs$-valued potential $\sa_{\mu}$ cf. \cite{Sperling:2019xar}
\small
\begin{align}
\label{graviton-h}
     h^{\mu\nu} &= \{y^\mu,\sa^\nu_{(1)}\}\big|_{0}+\{y^\nu,\sa^\mu_{(1)}\}\big|_{0} \nn\\
     &\approx\frac{\ell_p}{3}\Big[y^{\mu}\p_{\rho}\cA^{\nu|\rho}+y^{\nu}\p_{\rho}\cA^{\mu|\rho}-\cY \big(\cA^{\mu|\nu}+\cA^{\nu|\mu}\big)\Big]
     +\ell_p\big(\cA^{\mu|\nu}+\cA^{\nu|\mu}\big)
\end{align}
\normalsize
projecting along the fiber.
Since the terms in the first square bracket dominate, then to a very good approximation
\begin{align}
    h^{\mu\nu} \approx \frac{L_{\rm NC}^2}{R\cosh(\tau)}\Big[y^{\mu}\p_{\rho}\cA^{\nu|\rho}+y^{\nu}\p_{\rho}\cA^{\mu|\rho}-\cY \big(\cA^{\mu|\nu}+\cA^{\nu|\mu}\big)\Big]
\end{align}
in the late time regime. This illustrates the field redefinition relating the space-like and covariant field representations of the same $\hs$ degrees of freedom.
It is not hard to see that in the local $4d$ regime\footnote{Taking into account properly all conformal factors, the standard transformation law arises exactly \cite{Steinacker:2019awe}.} $h^{\mu\nu}$ also transforms covariantly under the linearized gauge transformations \cite{Steinacker:2019awe}:
\begin{align}\label{gra-gauge-transform}
    \delta_\L h_{\mu\nu} = \cL_\xi h_{\mu\nu} = 
 h_{\mu\nu} +  \del_\mu \xi_\nu \ +\del_\nu \xi_\mu 
\end{align}
where $\xi^\mu = \{\Lambda,y^\mu\}$ is a divergence-free vector field generated by $\Lambda$ \cite{Steinacker:2020xph}. The underlying gauge group consists of volume-preserving diffeomorphism in space-time, and $\hs$-valued generalizations thereof arising from symplectic vector fields on the underlying twistor space $\P^{1,2}$. 
This contains in particular all Lorentz transformations around any given point.
In that way, the appropriate action of Lorentz transformations arises for the tensor indices, which is not seen in the unitary formulation for $\hs$ modes. These observations strongly suggest that Lorentz invariance should effectively hold in the theory when we work with the covariant representation. This will be seen explicitly below for some simple examples.


Note that the inverse of the map \eqref{covariant-mapping} amounts to a non-local field redefinition, cf.  \cite{Steinacker:2019awe}. 
Therefore, the semi-classical HS-IKKT action {\em cannot} be written in terms of a local action using covariant fields, such as the covariant graviton \eqref{graviton-h}.
Rather, the appropriate kinetic terms for the covariant fields should arise at one loop where quantum effects dominate classical contributions in the spirit of Sakharov \cite{Sakharov:1967pk}.

This suggests the following interesting picture:
\medskip

\emph{Lorentzian HS-IKKT theory can be viewed either as local and unitary but mildly Lorentz-violating, or as Lorentz invariant but mildly non-local}.

\paragraph{$\hs$-valued gravitational coupling.} Using the above covariant graviton $h_{\mu\nu}$ in the local 4-dimensional regime, we can rewrite the vertex \eqref{main-cubic} in terms of a coupling between a $\hs$-valued graviton $h_{\mu\nu}$ and two $\hs$-valued scalar fields as follows: 
\begin{align}\label{covariant-cubic}
    \cV(\phi^{\ib}_{(s_1)},\phi_{\ib}^{(s_2)},\sa^{(s_3)})&=\int \mho\, \{\ttb^{\mu},\phi^{\ib}_{(s_1)}\}\{\phi_{\ib}^{(s_2)},\sa_{\mu}^{(s_3)}\}\approx e^{-\frac{3}{2}\tau}\int d^4y \,\p^{\mu}\phi^{\ib}_{(s_1)}\p^{\nu}\phi_{\ib}^{(s_2)}\{y_{\nu},\sa_{\mu}^{(s_3)}\}\nn\\
    &\approx e^{-\frac{3}{2}\tau}\int d^4y\,\partial_\mu\phi^{\ib}_{(s_1)}
    \partial_\nu\phi_{\ib}^{(s_2)}h^{\mu\nu}_{(s_3)} \, .
\end{align}
\normalsize
This is similar to the standard, minimal gravitational coupling between a covariant graviton and two scalar fields, cf. \eqref{3-vertices-002}. The vertex now looks perfectly covariant, even though all fields are $\hs$-valued. Notice also the absence of the non-commutativity scale $L_{\rm NC}$ in this formalism. 
Similarly, we can write the quartic vertex \eqref{quartic-vertex-aapp} as
    \begin{align}
    \cV(\phi^{\ib}_{(s_1)},\sa^{(s_2)},\phi_{\ib}^{(s_3)},\sa_{(s_4)})&=\int\mho\,\{\phi^{\ib}_{(s_1)},\sa^{\dot\mu}_{(s_2)}\},\{\phi_{\ib}^{(s_3)},\sa_{\dot\mu}^{(s_4)}\}\nn\\
    &\approx e^{-3\tau}\int d^4y \,\p^{\rho}\phi^{\ib}_{(s_1)}h_{\mu\rho}^{(s_2)} \p^{\sigma}\phi_{\ib}^{(s_3)}h^{\mu}{}_{\sigma (s_4)}\,.
\end{align}
Recall that the $\hs$-valued graviton $h_{\mu\nu}$ is not fundamental, but rather arises as {\em derivative} of the underlying space-like tensor field $\cA_{\mu|\nu(s)}(y)$.  
This is typical for Yang-Mills theories, where the physical fields are given as derivations of some potentials.
The same observation also applies to the full $\hs$-valued  fluctuation mode $\sa_{\mu}$ as shown above.
This illustrates how the strange-looking vertices in Section \ref{sec:spacelike-formulation} can be reconciled with the conventional, covariant language of QFT, and Lorentz invariance is recovered at least for some vertices.

Using 
\eqref{gra-gauge-transform}, it can be checked that the gravitational cubic vertex \eqref{covariant-cubic} obeys the standard Weinberg's soft theorem for low energy graviton. 
In addition, the 3-point amplitude of \eqref{covariant-cubic} reads
\begin{align}
    \cM_3\sim \delta^4(p_1+p_2+k_3)e^{-\frac{3}{2}\tau} (p_{1\mu}p_{2\nu}\varepsilon_3^{\mu\nu})+(1\leftrightarrow 2)\,,
\end{align}
which is non-trivial if the fields are would-be-massive modes.

\paragraph{$\hs$-valued Yang-Mills action.}

Finally, we briefly comment on an analogous covariant form of the cubic vertices \eqref{cubic-YM-vertex-U} for the gauge fields, and their quartic generalization.
These are part of the Yang-Mills term in the action, which does admit a covariant formulation along the lines of \cite{Steinacker:2010rh}.  
Viewing the $3+1$ semi-classical matrices $t^\mu = \ttb^\mu + \sa^\mu(y|u)$ as $\hs$-valued
functions on $\cM^{1,3}$ in the local coordinates $y^\mu$,
the Yang-Mills term of the action can be written in the local 4-dimensional regime as
$S_{\rm YM} \sim \int_{\cM^{1,3}} \, \cL_{\rm YM}$
 with
\begin{align}
 \cL_{\rm YM} := \{t^{\dot\mu},t^{\dot\nu}\}\{t_{\dot\mu},t_{\dot\nu}\} 
 =  \gamma^{\mu\nu} \gamma^{\rho\sigma}\sF_{\mu\rho}\sF_{\nu\sigma}=\rho^4G^{\mu\nu}G^{\rho\sigma}\sF_{\mu\rho}\sF_{\nu\sigma} \ . 
  \label{action-geometric-branes}
\end{align}
Here, $\sF_{\mu\rho}$ is a $\hs$-valued generalization of the symplectic form on $\cM^{1,3}$, and 
$\gamma^{\mu\nu}$ is the effective metric whose fluctuations are given by the graviton $h^{\mu\nu}$ \eqref{graviton-h}. 
All of these are tensorial objects transforming nicely under gauge transformations.
This strongly suggests that the associated interaction vertices should also transform covariantly under gauge transformation, and therefore respect Lorentz invariance in some form similar to the above.









\section{Discussion}\label{sec:discussion}
In this work, we studied 
tree-level scatterings on a local patch of FLRW cosmological spacetime $\cM^{1,3}$, aiming to understand the fate of 
the soft limit associated to these amplitudes. 
Our results show that Lorentzian HS-IKKT theory has non-trivial higher-spin scattering, due to the fact that the vertices
are mildly Lorentz- and parity-violating in the space-like formulation. 

To illustrate these results, we computed explicitly some 3-point scattering amplitudes in the space-like formulation of Lorentzian HS-IKKT. The amplitudes for the would-be-massive modes are non-trivial, but are strongly suppressed at low energies and in the weak curvature regime. In the massless sector, the amplitudes vanish on-shell. We also obtained the fusion rules for some cubic vertices with higher-spin fields in Lorentzian HS-IKKT, e.g. the couplings between two $\hs$-valued scalar fields with a $\hs$-valued gauge field. 
As an example, we study the simplest 4-point amplitude $\cM_4(0,1,1,0)$ with two scalars and two gauge fields $A_{\mu}$ on the external legs, which is strongly suppressed as expected.

To carry out these computations, we use two different ways of organizing the $\hs$ modes of the model: The first, ``space-like" or unitary representation makes the physical degrees of freedom very explicit, at the expense of manifest covariance. 
The second, ``covariant" formulation is closer to the standard formulation of Fronsdal-like higher-spin fields, but involves a non-trivial and non-local field redefinition. This formulation allows us to recast at least some space-like vertices in standard, covariant form. Then Lorentz invariance is effectively recovered at least for some sector of the theory, at the expense of manifest locality.

Our results provide further evidence that the higher-spin gauge theory induced by the IKKT matrix model on the present cosmological background is a consistent theory with physically reasonable and perhaps near-realistic features.



There are many interesting directions for future work on this model.
One obvious problem is to extend the present study to the non-abelian case, where higher-spin gauge fields take values in the $\msu(n)$ Lie algebra.
Another interesting project is to find a global spinorial formulation for Lorentzian HS-IKKT on $\cM^{1,3}$. In particular, the approach in \cite{Steinacker:2023zrb} based on spinors transforming under the compact space-like subgroup $SU(2)_L\times SU(2)_R\subset SU(2,2)$ is of limited use in the Lorentzian setting. To resolve this issue, one may try to consider the rotation subgroup to be the space-like isometry group $SL(2,\C)\simeq SO(1,3)$ where we can define Weyl spinors on $H^4$. 
This will be addressed in future work.

On the conceptual side, one important problem is to understand better the fate of Lorentz invariance, which is not manifest in the space-like/unitary formulation for higher-spin fields. 
Since the theory is invariant under $\hs$-valued 
volume-preserving diffeomorphism, Lorentz invariance is expected to hold effectively in some sense. We have seen that this can be made manifest for some fields and vertices in the covariant formulation, and a more general understanding of this issue would be very important.


Finally, as advertised in the introduction, a modified version of GR arises from quantum effects of the IKKT matrix model, 
as demonstrated in the one-loop effective action \cite{Steinacker:2021yxt}. The observed exponential suppression of the interaction vertices provides 
further justification for the weak coupling approach underlying that computation. Moreover, the present results suggest studying the loop amplitude of HS-IKKT theory using similar methods as in the present paper.

\section*{Acknowledgement}
We are grateful for useful discussions with Hikaru Kawai and Zhenya Skvortsov. We also thank an anonymous referee for various useful comments and suggestions for improving the draft. TT thanks the Erwin Schr\"odinger International Institute for Mathematics and Physics (ESI) for the hospitality during the workshop ``Large-N Matrix Models and Emergent Geometry'' in Vienna, where this work was initiated. This work is supported by the Austrian Science Fund (FWF) grant P36479.

\appendix

\section{Useful relations}\label{app:A}
This appendix provides additional relations for $\cM^{1,3}$ used in the main text. All the relations here are extracted from \cite{Sperling:2018xrm,Steinacker:2019awe,Sperling:2019xar}. Let $y_a$ for $a=0,1,2,3,4$ be coordinates of $H^4$ where $y_ay^a=-R^2$ and $y^a$ transform as vectors under $SO(1,4)$ with generator $m^{ab}$. After projecting out $y^4=R\sinh(\tau)$, we have the following relations
\begin{subequations}
\begin{align}
    y_{\mu}\ttb^{\mu}&=0\,,\qquad \qquad \qquad \qquad \quad \quad \qquad \qquad \qquad\quad \ \  \mu=0,1,2,3\,, \label{eq:orthogonalofPY}\\
    \eta_{\mu\nu}\ttb^{\mu}\ttb^{\nu}&=\frac{1}{\ell_p^2}+\frac{y_4^2}{\ell_p^2R^2}=+\ell_p^{-2}\cosh^2(\tau)\,, \qquad \qquad \quad \ \eta_{\mu\nu}=\diag(-,+,+,+)\,, \label{S2sphereM13}\\
   y^\mu y_\mu &= - R^2 \cosh^2(\tau) 
   \label{yy-square} \\ 
     \{\ttb^{\mu},y^{\nu}\}&=+\frac{\eta^{\mu\nu}}{R}y^4=\eta^{\mu\nu}\sinh(\tau)\,,\\
     \{\ttb^{\mu},\ttb^{\nu}\}&=-\frac{\theta^{\mu\nu}}{\ell_p^2R^2}=\frac{m^{\mu\nu}}{R^2}\,,\\
    \{\ttb^{\mu},y^4\}&=-\frac{y^{\mu}}{R}\,,\\
    m^{\mu\nu}&=-\frac{\theta^{\mu\nu}}{\ell_p^2}=-\frac{1}{\cosh^2(\tau)}\Big(\sinh(\tau)(y^{\mu}\ttb^{\nu}-y^{\nu}\ttb^{\mu})+\eps^{\mu\nu\sigma\rho}y_{\sigma}\ttb_{\rho}\Big)\label{mgenerator}\,,
    \end{align}
\end{subequations}
and
\begin{subequations}
    \begin{align}
    \ttb_{\mu}\theta^{\mu\nu}&=-\sinh(\tau)y^{\nu}\,,\label{eq:t-theta}\\
    y_{\mu}\theta^{\mu\nu}&=-\ell_p^2R^2\sinh(\tau)\ttb^{\mu}\label{eq:x-theta}\,,\\
    \theta_{\mu}{}^{\alpha}\theta^{\mu\beta}&=R^2\ell_p^2\eta^{\alpha\beta}-R^2\ell_p^4\,\ttb^{\alpha}\ttb^{\beta}+\ell_p^2\,y^{\alpha}y^{\beta}=R^2\ell_p^2\eta^{\alpha\beta}-R^2\ell_p^2\cosh^2(\tau)u^{\alpha}u^{\beta}+\ell_p^2\,y^{\alpha}y^{\beta}\label{eq:theta-theta}\,.
\end{align}
\end{subequations}
which are frequently used in the main text. Note that in the local physical regime
\begin{align}\label{flat-theta}
    \theta^{\mu\nu}\approx \frac{\ell_p\sinh(\tau)}{\cosh(\tau)}(y^{\mu}u^{\nu}-y^{\nu}u^{\mu})\,.
\end{align}

\paragraph{Projecting out fiber coordinates.}
Following \cite{Steinacker:2019awe}, we denote $[\cdot]_0$ as the projection to trivial harmonics which belong to the $\hs$-module $\cU^0$. Then,
\begin{align}
[u^{\mu_1}\ldots u^{\mu_{2s}}]_0=\alpha_{2s}\sum [u^{\mu_i}u^{\mu_j}]_0\ldots[u^{\mu_k} u^{\mu_l}]_0,\qquad \alpha_{2s}=\frac{2^s\times s!}{ (2s+1)!}\,,\quad s\geq 1\,
\end{align}
where
\begin{align}
    [u^{\mu}u^{\nu}]_0=\frac{1}{3}\Big[\eta^{\mu\nu}+\frac{x^{\mu}x^{\nu}}{R^2\cosh^2(\tau)}\Big]=:\hat\kappa^{\mu\nu}\,,\qquad \hat{\kappa}^{\mu\nu}x_{\mu}=0\,.
\end{align}
Note that in the large $R$ limit, all terms associated to the Levi-Civita symbols $\epsilon$ can be dropped. We also, have
\begin{subequations}
\begin{align}
    [\theta^{\mu\nu}\theta^{\rho\sigma}]_0&=\frac{\ell_p^2R^2}{3} \frac{\sinh^2(\tau)}{\cosh^2(\tau)} \Big(P^{\mu\sigma}P^{\nu\rho}-P^{\mu\rho}P^{\nu\sigma}+\frac{1}{R^2}\epsilon^{\mu\nu\rho\sigma}x_{\rho}\Big)\,,\\
    [\theta^{\mu_1\nu_1}\ldots \theta^{\mu_{2s}\nu_{2s}}]_0&=\alpha_{2s}\sum[\theta\theta]_0[\theta\theta]_0\ldots[\theta\theta]_0 \,,\\
     [\theta^{a_1b_1}\ldots \theta^{a_{2s+1}b_{2s+1}}]_0&=0\,,
\end{align}
\end{subequations}
where now $P^{\mu\nu}=\eta^{\mu\nu}+\frac{x^{\mu}x^{\nu}}{R^2}$. Besides the above, we will also sometime use
\begin{subequations}
    \begin{align}
    [\theta^{\mu\nu}]_0&=0\,,\\
    [\theta^{\mu\nu}u^{\rho}]_0&=\frac{\ell_p\sinh(\tau)}{3\cosh(\tau)}\Big([\eta^{\rho\nu}x^{\mu}-\eta^{\rho\mu}x^{\nu}]+\frac{x_{\beta}\epsilon^{\beta \rho\mu\nu}}{\sinh(\tau)}\Big)\,,\\
     [\theta^{\alpha\beta}u^{\mu_1}\ldots u^{\mu_{2s+1}}]_0&=\alpha_{2s}\sum_{i=1}^{2s+1}[\theta^{\alpha\beta}u^{\mu_i}]_0[uu]_0\ldots[uu]_0\,,\\
     [\theta^{\alpha\beta}u^{\mu_1}\ldots u^{\mu_{2s}}]_0&=0\,,
\end{align}
\end{subequations}
in the main text. 
\section{Derivation of the IR mass}\label{app:IRmass}
This appendix derives the IR mass resulting from acting $\Box_{1,3}$ on the polynomials $u^{\mu(s)}$.\footnote{This also corrects the corresponding formula (5.4.65) in \cite{Steinacker:2024unq}.}  To simplify the computations, we note the following useful relations:
\small
    \begin{align}
    \{\ttb^{\mu},\ttb^{\nu}\}&=-\frac{\theta^{\mu\nu}}{\ell_p^2R^2}\,,\quad \{\ttb^{\mu},\frac{1}{\cosh(\tau)}\}\approx\frac{y^{\mu}}{R^2\cosh^2(\tau)}\,,\quad \{\ttb^{\mu},\frac{1}{\cosh^2(\tau)}\}\approx \frac{2y^{\mu}}{R^2\cosh^3(\tau)}\,.
\end{align}
\normalsize
Then,
\begin{align}
     \{\ttb^{\mu},u^{\nu}\}=-\frac{\theta^{\mu\nu}}{\cosh(\tau)\ell_pR^2}+\frac{\ell_p \ttb^{\nu}y^{\mu}}{R^2\cosh^2(\tau)}\,.
\end{align}
Next, we can compute
\begin{align}
    \{\ttb^{\dot\mu},u^{\nu}\}\{\ttb_{\dot\mu},u^{\nu}\}&=\Big(-\frac{\theta^{\mu\nu}}{\cosh(\tau)\ell_pR^2}+\frac{\ell_p \ttb^{\nu}y^{\mu}}{R^2\cosh^2(\tau)}\Big)\Big(-\frac{\theta_{\mu}{}^{\nu}}{\cosh(\tau)\ell_pR^2}+\frac{\ell_p \ttb^{\nu}y_{\mu}}{R^2\cosh^2(\tau)}\Big)\nn\\
    &=\frac{\theta^{\mu\nu}\theta_{\mu}{}^{\nu}}{\cosh^2(\tau)\ell_p^2R^4}+\frac{\ell_p^2\ttb^{\nu(2)}y^2}{R^4\cosh^4(\tau)}-2\frac{\ttb^{\nu}y^{\mu}\theta_{\mu}{}^{\nu}}{R^4\cosh^3(\tau)}\nn\\
    &\approx\Big(-\frac{\ell_p^2}{R^2\cosh^2(\tau)}-\frac{\ell_p^2}{R^2\cosh^2(\tau)}+2\frac{\ell_p^2}{R^2\cosh^2(\tau)}\Big)\ttb^{\nu(2)}\nn\\
    &\approx 0\,.
\end{align}
Hence, there are no contributions that will be quadratic in spins. The last step of the computation is to act $\{\ttb^{\dot\mu},\{\ttb_{\dot\mu},.\}\}$ on $u^{\nu}$. We obtain
\begin{align}
    s\{\ttb^{\dot\mu},\{\ttb_{\dot\mu},u^{\nu}\}u^{\nu(s-1)}\}
    &=s(s-1)\{\ttb^{\dot\mu},u^{\nu}\}\{\ttb_{\dot\mu},u^{\nu}\}u^{\nu(s-2)}+s\{\ttb^{\dot\mu},\{\ttb_{\dot\mu},u^{\nu}\}\}u^{\nu(s-1)}\,.
\end{align}
\normalsize
It is now a simple computation to show that
\begin{align}
    \{\ttb^{\dot\mu},\{\ttb_{\dot\mu},u^{\nu}\}\}&=\ell_p\{\ttb^{\dot\mu},\{\ttb_{\dot\mu},\frac{\ttb^{\nu}}{\cosh(\tau)}\}\}\nn\\
    &=-\frac{3\ell_p\ttb^{\nu}}{R^2\cosh(\tau)}
    +\ell_p\ttb^{\nu}\{\ttb^{\dot\mu},\{\ttb_{\dot\mu},\frac{1}{\cosh(\tau)}\}\}
    +2\ell_p\{\ttb^{\dot\mu},\ttb^{\nu}\}\{\ttb_{\dot\mu},\frac{1}{\cosh(\tau)}\}\nn\\
    &=-\frac{3u^{\nu}}{R^2}+\frac{\ell_p\ttb^{\nu}}{R^2}\{\ttb^{\dot\mu},\frac{y_{\dot\mu}}{\cosh^2(\tau)}\}-\frac{2}{\ell_pR^2}\frac{\theta^{\dot\mu\nu}y_{\dot\mu}}{R^2\cosh^2(\tau)}\nn\\
    &\approx-\frac{3 u^{\nu}}{R^2}+\frac{4\ell_p\ttb^{\nu}}{R^2\cosh(\tau)}+\frac{2\ell_p\ttb^{\nu}y^2}{R^4\cosh^3(\tau)}+\frac{2\ell_p\ttb^{\nu}}{R^2\cosh(\tau)}\nn\\
    &\approx +\frac{1}{R^2}u^{\nu}\,,
\end{align}
using 
\begin{align}
    \{\theta^{\mu\nu},\ttb^{\sigma}\}=-\ell_p^2\Big(\eta^{\nu\sigma}\ttb^{\mu}-\eta^{\mu\sigma}\ttb^{\nu}\Big)\,.
\end{align}
Thus,
\begin{align}
    \Box_{1,3}u^{\nu(s)}=-\{\ttb^{\dot\mu},\{\ttb_{\dot\mu},u^{\nu(s)}\}\}=-\frac{s}{R^2}u^{\nu(s)}\,.
\end{align}

\section{Vertices at \texorpdfstring{$\Lambda=4$}{L4}}\label{app:B}
This appendix derives all sub-vertices of \eqref{main-cubic} at $\Lambda=4$, where
\begin{align}
    \cV_3^{[4]}=&+\cV^{(0,0,4)}_3+\cV_3^{(1,0,3)}+\cV_3^{(0,1,3)}+\cV_3^{(0,2,2)}+\cV_3^{(2,0,2)}+\cV_3^{(1,1,2)}\nn\\
    &+\cV_3^{(2,1,1)}+\cV_3^{(1,2,1)}+\cV_3^{(3,0,1)}+\cV_3^{(0,3,1)}\,.
\end{align}
\normalsize
Below, the pre-factors of all sub-vertices $\cV_3^{(s_1,s_2,s_3)}$ is $\frac{2\ell_p}{27\cosh(\tau)\sqrt{\sinh(\tau)}}$. We will suppress this factor to simplify our expressions.

Firstly, due to tracelessness 
\begin{align*}
    \cV_3^{(0,0,4)}&\sim 0\,,\\
    \widetilde\cV_3^{(0,0,2)}&\sim 0\,.
\end{align*}
Next, we have
\footnotesize
\begin{align*}
    \cV_3^{(1,0,3)}&\sim \int \Big[(\cY\varphi_{\ib}^{\alpha_3})(\p^{\alpha_2}\p^{\mu}\phi^{\ib})-(\p^{\alpha_2}\varphi_{\ib}^{\alpha_3})(\cY\p^{\mu}\phi^{\ib})\Big]\cA_{\mu\alpha_2\alpha_3}+\frac{y_{\beta}\epsilon^{\beta\alpha_2\rho\sigma}}{\sinh(\tau)}(\p_{\rho}\varphi_{\ib}^{\alpha_3})(\p_{\sigma}\p^{\mu}\phi^{\ib})\cA_{\mu\alpha_2\alpha_3}\,,\\
    \cV_3^{(0,1,3)}&\sim \int \Big[(\cY\phi_{\ib})(\p^{\alpha_2}\p^{\mu}\varphi^{\ib\alpha_3})-(\p^{\alpha_2}\phi_{\ib})(\cY \p^{\mu}\varphi^{\ib\alpha_3})\Big]\cA_{\mu\alpha_2\alpha_3}+\frac{y_{\beta}\epsilon^{\beta\alpha_2\rho\sigma}}{\sinh(\tau)}(\p_{\rho}\phi_{\ib})(\p_{\sigma}\p^{\mu}\varphi^{\ib \alpha_3})\cA_{\mu\alpha_2\alpha_3}\,.
\end{align*}
\normalsize
\normalsize
Similarly, we get
\footnotesize
\begin{align*}
    \cV_3^{(0,2,2)}&\sim \int \Big[(\cY\phi_{\ib})(\p^{\alpha_1}\p^{\mu}\varphi^{\ib}_{\alpha_1\alpha_2})-(\p^{\alpha_1}\phi_{\ib})(\cY \p^{\mu}\varphi^{\ib}_{\alpha_1\alpha_2})\Big]\cA_{\mu}{}^{\alpha_2}+\frac{y_{\beta}\epsilon^{\beta\alpha_1\rho\sigma}}{\sinh(\tau)}(\p_{\rho}\phi_{\ib})(\p_{\sigma}\p^{\mu}\varphi^{\ib}_{\alpha_1\alpha_2})\cA_{\mu}{}^{\alpha_2}\,,\\
    \cV_3^{(2,0,2)}&\sim \int \Big[(\cY \varphi_{\ib\alpha_1\alpha_2})(\p^{\alpha_1}\p^{\mu}\phi^{\ib})-(\p^{\alpha_1}\varphi_{\ib\alpha_1\alpha_2})(\cY \p^{\mu}\phi^{\ib})\Big]\cA_{\mu}{}^{\alpha_2}+\frac{y_{\beta}\epsilon^{\beta\alpha_1\rho\sigma}}{\sinh(\tau)}(\p_{\rho}\varphi_{\ib\alpha_1\alpha_2})(\p_{\sigma}\p^{\mu}\phi^{\ib})\cA_{\mu}{}^{\alpha_2}\,,\\
    \cV_3^{(1,1,2)}&\sim \frac{1}{2}\Bigg(\int \Big[(\cY\varphi_{\ib\alpha_1})(\p^{\alpha_1}\p^{\mu}\varphi^{\ib}_{\alpha_2})-(\p^{\alpha_1}\varphi_{\alpha_1})(\cY \p^{\mu}\varphi^{\ib}_{\alpha_2})\Big]\cA_{\mu}^{\alpha_2}+\frac{y_{\beta}\epsilon^{\beta\alpha_1\rho\sigma}}{\sinh(\tau)}(\p_{\rho}\varphi_{\ib \alpha_1})(\p_{\sigma}\p^{\mu}\varphi^{\ib}_{\alpha_2})\cA_{\mu}^{\alpha_2}\\
    &\quad +\int \Big[(\cY\varphi_{\ib\alpha_1})(\p^{\alpha_2}\p^{\mu}\varphi^{\ib}_{\alpha_2})-(\p^{\alpha_2}\varphi_{\ib\alpha_1})(\cY \p^{\mu}\varphi^{\ib}_{\alpha_2})\Big]\cA_{\mu}^{\alpha_1}+\frac{y_{\beta}\epsilon^{\beta\alpha_2\rho\sigma}}{\sinh(\tau)}(\p_{\rho}\varphi_{\ib \alpha_1})(\p_{\sigma}\p^{\mu}\varphi^{\ib}_{\alpha_2})\cA_{\mu}^{\alpha_1}\\
    &\quad +\int \Big[(\cY \varphi_{\ib \alpha_1})(\p^{\alpha_3}\p^{\mu}\varphi^{\ib\alpha_1})-(\p^{\alpha_3}\varphi_{\ib\alpha_1})(\cY \p^{\mu}\varphi^{\ib\alpha_1})\Big]\cA_{\mu\alpha_3}+\frac{y_{\beta}\epsilon^{\beta\alpha_3\rho\sigma}}{\sinh(\tau)}(\p_{\rho}\varphi_{\ib\alpha_1})(\p_{\sigma}\p^{\mu}\varphi^{\ib\alpha_1})\cA_{\mu\alpha_3} \Bigg)\,.
\end{align*}
\normalsize
Finally, we consider $\cV^{(s_1,s_2,1)}$ sub-vertices
\footnotesize
\begin{align*}
    \cV_3^{(3,0,1)}&\sim 0\\
    \cV_3^{(0,3,1)}&\sim 0\\
    \cV_3^{(2,1,1)}&\sim \int \Big[(\cY \varphi_{\ib\alpha_1\alpha_2})(\p^{\alpha_1}\p^{\mu}\varphi^{\ib\alpha_2})-(\p^{\alpha_1}\varphi_{\ib\alpha_1\alpha_2})(\cY \p^{\mu}\varphi^{\ib\alpha_2})\Big]\cA_{\mu}+\frac{y_{\beta}\epsilon^{\beta\alpha_1\rho\sigma}}{\sinh(\tau)}(\p_{\rho}\varphi_{\ib\alpha_1\alpha_2})(\p_{\sigma}\p^{\mu}\varphi^{\ib\alpha_2})\cA_{\mu}\\
    \cV_3^{(1,2,1)}&\sim \int \Big[(\cY \varphi_{\ib}^{\alpha_3})(\p^{\alpha_2}\p^{\mu}\varphi^{\ib}_{\alpha_2\alpha_3})-(\p^{\alpha_2}\varphi_{\ib}^{\alpha_3})(\cY \p^{\mu}\varphi^{\ib}_{\alpha_2\alpha_3})\Big]\cA_{\mu}+\frac{y_{\beta}\epsilon^{\beta\alpha_2\rho\sigma}}{\sinh(\tau)}(\p_{\rho}\varphi_{\ib}^{\alpha_3})(\p_{\sigma}\p^{\mu}\varphi^{\ib}_{\alpha_2\alpha_3})\cA_{\mu}\,.
\end{align*}
\normalsize
Note that $\cV_3^{(0,s_2\geq 3,1)=0}$ and $\cV_3^{(s_1\geq 3,0,1)}=0$ due to tracelessness. All vertices for other higher-spin cases can be obtained similarly.

\section{On higher-spin conserved currents and Noether couplings}\label{app:noether}
 Let us explore a more ``standard'' non-commutative version of higher-spin currents coupled to higher-spin fields. Recall that in field theory, conserved higher-spin currents are defined as $J_s=\phi^*(\overleftrightarrow{\p})^s\phi$. Thus, in a non-commutative setting, we may consider
\begin{align}
    J_{\mu}&=\phi^*\{\ttb_{\mu},\phi\}-\{\ttb_{\mu},\phi^*\}\phi\,,\\
    J_{\mu\nu}&=\phi^*\{\ttb_{(\mu},\{\ttb_{\nu)},\phi\}\}+\{\ttb_{(\mu},\{t_{\nu)},\phi^*\}\}\phi-\{\ttb_{\mu},\phi^*\}\{\ttb_{\nu},\phi\}-\{\ttb_{\nu},\phi^*\}\{\ttb_{\mu},\phi\}\,,
\end{align}
where $\phi,\phi^*$ are some complex scalar fields. 
Given the on-shell relation $\Box\phi:=\{\ttb^{\mu},\{\ttb_{\mu},\phi\}\}=0$ in the local $4d$ regime, we obtain
\begin{align}
    \{\ttb^{\mu},J_{\mu}\}&\approx 0\,,\\
    \{\ttb^{\mu},J_{\mu\nu}\}&= \{\ttb^{\mu},\{\ttb_{\mu},\{\ttb_{\nu},\phi^*\}\}\}\phi+\phi^*\{\ttb^{\mu},\{\ttb_{\mu},\{\ttb_{\nu},\phi\}\}\}\approx0
\end{align}
where 
\begin{align}
    \{\ttb^{\mu},\{\ttb_{\mu},\{\ttb_{\nu},\phi\}\}\}&=\{\ttb_{\nu},\{\ttb^{\mu},\{\ttb_{\mu},\phi\}\}\}+\{\{\ttb_{\nu},\ttb^{\mu}\},\{\ttb_{\mu},\phi\}\}+\{\ttb^{\mu},\{\{\ttb_{\mu},\ttb_{\nu}\},\phi\}\}\nn\\
    &=\{\ttb_{\nu},\Box \phi\}-\{\theta^{\mu\nu},\{\ttb_{\mu},\phi\}\}+\{\ttb^{\mu},\{\theta_{\mu\nu},\phi\}\}\nn\\
    &=\{\ttb_{\nu},\Box \phi\}\,.
\end{align}
Here, $\{\ttb_{\mu},\{\ttb_{\nu},-\}\}=\{\ttb_{\nu},\{\ttb_{\mu},-\}\}+\{\theta_{\mu\nu},-\}$, so the currents will be conserved on-shell only in the IR regime where we can set $\{\theta^{\mu\nu},-\}\sim 0$ as $\theta^{\mu\nu}$ can be treated as constant in the flat limit. Note that when we couple conserved currents to symmetric higher-spin gauge fields, all the anti-symmetric components vanish due to symmetry.

Armed with the above currents, let us explore two cases that are often considered in the literature:
\begin{itemize}
    \item[-] \underline{Minimal coupling to higher-spin fields:} Consider the following Noether couplings:
\begin{align}\label{Noether1}
    V_s^{\text{(m)}}=\sum_s\tg^s\int  J_{\mu(s)}A^{\mu(s)}\,,\qquad \delta A^{\mu(s)}=\{\ttb^\mu,\xi^{\mu(s-1)}\}=\sinh(\tau)\p^{\mu}\xi^{\mu(s-1)}\,.
\end{align}
where $J_{\mu(s)}$ is a suitable generalization of the above currents. This type of vertices can be trivially evaluated by noting that $\theta_{\mu\nu}A^{\mu\nu\rho(s-2)}=0$. Then, as in Weinberg \cite{Weinberg:1964ew}, gauge invariance of the $S$-matrix would imply 
\begin{align}\label{Weinberg-constraint}
    \sum_{i=1}^n \tg_i^s\,p_i^{\mu_1}\ldots p_i^{\mu_{s-1}}=0\,,
\end{align}
where $p_i^{\mu}$ are external momenta of the scalar field $\phi_i$. Thus, for $s\geq 3$, the only solution to \eqref{Weinberg-constraint} is to set $\tg_i^s=0$. In other words, there cannot be higher-spin interactions in the form of \eqref{Noether1} in such a scenario. 
Therefore such a ``Noether construction” does not seem useful in the present setting, where the matrix model provides directly a gauge-invariant action.

    \item[-] \underline{Non-minimal coupling to higher-spin fields:} Consider for instance
\begin{align}
    V_s^{\text{(n.m)}}=\int \{\ttb_{(\mu_1},\{\ldots,\{\ttb_{\mu_{s-i}},J_{\mu_{s-i+1}\ldots \mu_{s})}\}\ldots\}\}A^{\mu_1\ldots \mu_s}\,.
\end{align}
By integrating by part, we can rewrite the above as 
\begin{align}
    V_s^{\text{(n.m)}}\sim \int  J_{(\mu_{s-i+1}\ldots \mu_s}\{\ttb_{\mu_{s-i}},\{\ldots,\{\ttb_{\mu_1)},A^{\mu_1\ldots \mu_s}\}\ldots\}\}\,.
\end{align}
Hence in the gauge where $\{\ttb_{\mu},A^{\mu \nu(s-1)}\}=0$, the non-minimal couplings vanishes. Note that this gauge can always be found, as shown in \cite{Steinacker:2019awe}.\footnote{See also \cite{Taylor:1999gq} for other non-minimal couplings between a graviton and higher-spin extension of the stress-energy tensor.}

\end{itemize}

The above examples indicate that local Noether-type vertices that are parity- and Lorentz-invariant cannot couple to macroscopic higher-spin fields. In contrast, the vertices of Lorentzian HS-IKKT lead to non-trivial scattering amplitudes. This provides the first example of a local unitary higher-spin theory in Lorentzian signature that can scatter in flat spacetime.

\setstretch{0.8}
\footnotesize
\bibliography{twistor}

\providecommand{\href}[2]{#2}\begingroup\raggedright\begin{thebibliography}{10}

\bibitem{Sakharov:1967pk}
A.~D. Sakharov, {\it {Vacuum quantum fluctuations in curved space and the
  theory of gravitation}},  {\em Dokl. Akad. Nauk Ser. Fiz.} {\bf 177} (1967)
  70--71.

\bibitem{Visser:2002ew}
M.~Visser, {\it {Sakharov's induced gravity: A Modern perspective}},  {\em Mod.
  Phys. Lett. A} {\bf 17} (2002) 977--992
  [\href{http://arXiv.org/abs/gr-qc/0204062}{{\tt gr-qc/0204062}}].

\bibitem{Ishibashi:1996xs}
N.~Ishibashi, H.~Kawai, Y.~Kitazawa and A.~Tsuchiya, {\it {A Large N reduced
  model as superstring}},  {\em Nucl. Phys. B} {\bf 498} (1997) 467--491
  [\href{http://arXiv.org/abs/hep-th/9612115}{{\tt hep-th/9612115}}].

\bibitem{Steinacker:2019fcb}
H.~C. Steinacker, {\it {On the quantum structure of space-time, gravity, and
  higher spin in matrix models}},  {\em Class. Quant. Grav.} {\bf 37} (2020),
  no.~11 113001 [\href{http://arXiv.org/abs/1911.03162}{{\tt 1911.03162}}].

\bibitem{Steinacker:2021yxt}
H.~C. Steinacker, {\it {Gravity as a quantum effect on quantum space-time}},
  {\em Phys. Lett. B} {\bf 827} (2022) 136946
  [\href{http://arXiv.org/abs/2110.03936}{{\tt 2110.03936}}].

\bibitem{Steinacker:2023myp}
H.~C. Steinacker, {\it {One-loop effective action and emergent gravity on
  quantum spaces in the IKKT matrix model}},
  \href{http://arXiv.org/abs/2303.08012}{{\tt 2303.08012}}.

\bibitem{Anagnostopoulos:2022dak}
K.~N. Anagnostopoulos, T.~Azuma, K.~Hatakeyama, M.~Hirasawa, Y.~Ito,
  J.~Nishimura, S.~K. Papadoudis and A.~Tsuchiya, {\it {Progress in the
  numerical studies of the type IIB matrix model}},  10, 2022.
\newblock \href{http://arXiv.org/abs/2210.17537}{{\tt 2210.17537}}.

\bibitem{Nishimura:2019qal}
J.~Nishimura and A.~Tsuchiya, {\it {Complex Langevin analysis of the space-time
  structure in the Lorentzian type IIB matrix model}},  {\em JHEP} {\bf 06}
  (2019) 077 [\href{http://arXiv.org/abs/1904.05919}{{\tt 1904.05919}}].

\bibitem{Sperling:2018xrm}
M.~Sperling and H.~C. Steinacker, {\it {The fuzzy 4-hyperboloid $H^4_n$ and
  higher-spin in Yang\textendash{}Mills matrix models}},  {\em Nucl. Phys. B}
  {\bf 941} (2019) 680--743 [\href{http://arXiv.org/abs/1806.05907}{{\tt
  1806.05907}}].

\bibitem{Steinacker:2023zrb}
H.~Steinacker and T.~Tran, {\it {Spinorial higher-spin gauge theory from IKKT
  in Euclidean and Minkowski signatures}},
  \href{http://arXiv.org/abs/2305.19351}{{\tt 2305.19351}}.

\bibitem{Fradkin:1986ka}
E.~S. Fradkin and M.~A. Vasiliev, {\it {Candidate to the Role of Higher Spin
  Symmetry}},  {\em Annals Phys.} {\bf 177} (1987) 63.

\bibitem{Steinacker:2019awe}
H.~C. Steinacker, {\it {Higher-spin kinematics \& no ghosts on quantum
  space-time in Yang-Mills matrix models}},
  \href{http://arXiv.org/abs/1910.00839}{{\tt 1910.00839}}.

\bibitem{Sperling:2019xar}
M.~Sperling and H.~C. Steinacker, {\it {Covariant cosmological quantum
  space-time, higher-spin and gravity in the IKKT matrix model}},  {\em JHEP}
  {\bf 07} (2019) 010 [\href{http://arXiv.org/abs/1901.03522}{{\tt
  1901.03522}}].

\bibitem{Weinberg:1964ew}
S.~Weinberg, {\it {Photons and Gravitons in $S$-Matrix Theory: Derivation of
  Charge Conservation and Equality of Gravitational and Inertial Mass}},  {\em
  Phys. Rev.} {\bf 135} (1964) B1049--B1056.

\bibitem{Coleman:1967ad}
S.~R. Coleman and J.~Mandula, {\it {All Possible Symmetries of the S Matrix}},
  {\em Phys. Rev.} {\bf 159} (1967) 1251--1256.

\bibitem{Steinacker:2017vqw}
H.~C. Steinacker, {\it {Cosmological space-times with resolved Big Bang in
  Yang-Mills matrix models}},  {\em JHEP} {\bf 02} (2018) 033
  [\href{http://arXiv.org/abs/1709.10480}{{\tt 1709.10480}}].

\bibitem{Steinacker:2022yhs}
H.~C. Steinacker, {\it {Classical space-time geometry and the weak gravity
  regime in the IKKT matrix model}},  {\em PoS} {\bf CORFU2021} (2022) 232
  [\href{http://arXiv.org/abs/2204.05679}{{\tt 2204.05679}}].

\bibitem{Casali:2014xpa}
E.~Casali, {\it {Soft sub-leading divergences in Yang-Mills amplitudes}},  {\em
  JHEP} {\bf 08} (2014) 077 [\href{http://arXiv.org/abs/1404.5551}{{\tt
  1404.5551}}].

\bibitem{Cachazo:2014fwa}
F.~Cachazo and A.~Strominger, {\it {Evidence for a New Soft Graviton Theorem}},
   \href{http://arXiv.org/abs/1404.4091}{{\tt 1404.4091}}.

\bibitem{Bern:2014vva}
Z.~Bern, S.~Davies, P.~Di~Vecchia and J.~Nohle, {\it {Low-Energy Behavior of
  Gluons and Gravitons from Gauge Invariance}},  {\em Phys. Rev. D} {\bf 90}
  (2014), no.~8 084035 [\href{http://arXiv.org/abs/1406.6987}{{\tt
  1406.6987}}].

\bibitem{Hamada:2018vrw}
Y.~Hamada and G.~Shiu, {\it {Infinite Set of Soft Theorems in Gauge-Gravity
  Theories as Ward-Takahashi Identities}},  {\em Phys. Rev. Lett.} {\bf 120}
  (2018), no.~20 201601 [\href{http://arXiv.org/abs/1801.05528}{{\tt
  1801.05528}}].

\bibitem{Campoleoni:2017mbt}
A.~Campoleoni, D.~Francia and C.~Heissenberg, {\it {On higher-spin
  supertranslations and superrotations}},  {\em JHEP} {\bf 05} (2017) 120
  [\href{http://arXiv.org/abs/1703.01351}{{\tt 1703.01351}}].

\bibitem{Miller:2022fvc}
N.~Miller, A.~Strominger, A.~Tropper and T.~Wang, {\it {Soft Gravitons in the
  BFSS Matrix Model}},  \href{http://arXiv.org/abs/2208.14547}{{\tt
  2208.14547}}.

\bibitem{Low:1954kd}
F.~E. Low, {\it {Scattering of light of very low frequency by systems of spin
  1/2}},  {\em Phys. Rev.} {\bf 96} (1954) 1428--1432.

\bibitem{Gell-Mann:1954wra}
M.~Gell-Mann and M.~L. Goldberger, {\it {Scattering of low-energy photons by
  particles of spin 1/2}},  {\em Phys. Rev.} {\bf 96} (1954) 1433--1438.

\bibitem{Low:1958sn}
F.~E. Low, {\it {Bremsstrahlung of very low-energy quanta in elementary
  particle collisions}},  {\em Phys. Rev.} {\bf 110} (1958) 974--977.

\bibitem{kazes1959generalized}
E.~Kazes, {\it Generalized current conservation and low energy limit of photon
  interactions},  {\em Il Nuovo Cimento (1955-1965)} {\bf 13} (1959)
  1226--1239.

\bibitem{Gross:1968in}
D.~J. Gross and R.~Jackiw, {\it {Low-Energy Theorem for Graviton Scattering}},
  {\em Phys. Rev.} {\bf 166} (1968) 1287--1292.

\bibitem{jackiw1968low}
R.~Jackiw, {\it Low-energy theorems for massless bosons: photons and
  gravitons},  {\em Physical Review} {\bf 168} (1968), no.~5 1623.

\bibitem{Saito:1969lga}
S.~Saito, {\it {Low-energy theorem for Compton scattering}},  {\em Phys. Rev.}
  {\bf 184} (1969) 1894--1902.

\bibitem{Ferrari:1971at}
R.~Ferrari and L.~E. Picasso, {\it {Spontaneous breakdown in quantum
  electrodynamics}},  {\em Nucl. Phys. B} {\bf 31} (1971) 316--330.

\bibitem{Tran:2022amg}
T.~Tran, {\it {Constraining higher-spin S-matrices}},
  \href{http://arXiv.org/abs/2212.02540}{{\tt 2212.02540}}.

\bibitem{Tseytlin:2002gz}
A.~A. Tseytlin, {\it {On limits of superstring in AdS(5) x S**5}},  {\em Theor.
  Math. Phys.} {\bf 133} (2002) 1376--1389
  [\href{http://arXiv.org/abs/hep-th/0201112}{{\tt hep-th/0201112}}].

\bibitem{Segal:2002gd}
A.~Y. Segal, {\it {Conformal higher spin theory}},  {\em Nucl. Phys. B} {\bf
  664} (2003) 59--130 [\href{http://arXiv.org/abs/hep-th/0207212}{{\tt
  hep-th/0207212}}].

\bibitem{Bekaert:2010ky}
X.~Bekaert, E.~Joung and J.~Mourad, {\it {Effective action in a higher-spin
  background}},  {\em JHEP} {\bf 02} (2011) 048
  [\href{http://arXiv.org/abs/1012.2103}{{\tt 1012.2103}}].

\bibitem{Basile:2022nou}
T.~Basile, M.~Grigoriev and E.~Skvortsov, {\it {Covariant action for conformal
  higher spin gravity}},  {\em J. Phys. A} {\bf 56} (2023), no.~38 385402
  [\href{http://arXiv.org/abs/2212.10336}{{\tt 2212.10336}}].

\bibitem{Metsaev:1991mt}
R.~R. Metsaev, {\it {Poincare invariant dynamics of massless higher spins:
  Fourth order analysis on mass shell}},  {\em Mod. Phys. Lett. A} {\bf 6}
  (1991) 359--367.

\bibitem{Metsaev:1991nb}
R.~R. Metsaev, {\it {S matrix approach to massless higher spins theory. 2: The
  Case of internal symmetry}},  {\em Mod. Phys. Lett. A} {\bf 6} (1991)
  2411--2421.

\bibitem{Ponomarev:2016lrm}
D.~Ponomarev and E.~D. Skvortsov, {\it {Light-Front Higher-Spin Theories in
  Flat Space}},  {\em J. Phys. A} {\bf 50} (2017), no.~9 095401
  [\href{http://arXiv.org/abs/1609.04655}{{\tt 1609.04655}}].

\bibitem{Ponomarev:2017nrr}
D.~Ponomarev, {\it {Chiral Higher Spin Theories and Self-Duality}},  {\em JHEP}
  {\bf 12} (2017) 141 [\href{http://arXiv.org/abs/1710.00270}{{\tt
  1710.00270}}].

\bibitem{Metsaev:2018xip}
R.~R. Metsaev, {\it {Light-cone gauge cubic interaction vertices for massless
  fields in AdS(4)}},  {\em Nucl. Phys. B} {\bf 936} (2018) 320--351
  [\href{http://arXiv.org/abs/1807.07542}{{\tt 1807.07542}}].

\bibitem{Metsaev:2019dqt}
R.~R. Metsaev, {\it {Cubic interaction vertices for N=1 arbitrary spin massless
  supermultiplets in flat space}},  {\em JHEP} {\bf 08} (2019) 130
  [\href{http://arXiv.org/abs/1905.11357}{{\tt 1905.11357}}].

\bibitem{Metsaev:2019aig}
R.~R. Metsaev, {\it {Cubic interactions for arbitrary spin $ \mathcal{N} $
  -extended massless supermultiplets in 4d flat space}},  {\em JHEP} {\bf 11}
  (2019) 084 [\href{http://arXiv.org/abs/1909.05241}{{\tt 1909.05241}}].

\bibitem{Tsulaia:2022csz}
M.~Tsulaia and D.~Weissman, {\it {Supersymmetric Quantum Chiral Higher Spin
  Gravity}},  \href{http://arXiv.org/abs/2209.13907}{{\tt 2209.13907}}.

\bibitem{Krasnov:2021nsq}
K.~Krasnov, E.~Skvortsov and T.~Tran, {\it {Actions for self-dual Higher Spin
  Gravities}},  {\em JHEP} {\bf 08} (2021) 076
  [\href{http://arXiv.org/abs/2105.12782}{{\tt 2105.12782}}].

\bibitem{Adamo:2022lah}
T.~Adamo and T.~Tran, {\it {Higher-spin Yang-Mills, amplitudes and
  self-duality}},  \href{http://arXiv.org/abs/2210.07130}{{\tt 2210.07130}}.

\bibitem{Joung:2015eny}
E.~Joung, S.~Nakach and A.~A. Tseytlin, {\it {Scalar scattering via conformal
  higher spin exchange}},  {\em JHEP} {\bf 02} (2016) 125
  [\href{http://arXiv.org/abs/1512.08896}{{\tt 1512.08896}}].

\bibitem{Beccaria:2016syk}
M.~Beccaria, S.~Nakach and A.~A. Tseytlin, {\it {On triviality of S-matrix in
  conformal higher spin theory}},  {\em JHEP} {\bf 09} (2016) 034
  [\href{http://arXiv.org/abs/1607.06379}{{\tt 1607.06379}}].

\bibitem{Roiban:2017iqg}
R.~Roiban and A.~A. Tseytlin, {\it {On four-point interactions in massless
  higher spin theory in flat space}},  {\em JHEP} {\bf 04} (2017) 139
  [\href{http://arXiv.org/abs/1701.05773}{{\tt 1701.05773}}].

\bibitem{Skvortsov:2018jea}
E.~D. Skvortsov, T.~Tran and M.~Tsulaia, {\it {Quantum Chiral Higher Spin
  Gravity}},  {\em Phys. Rev. Lett.} {\bf 121} (2018), no.~3 031601
  [\href{http://arXiv.org/abs/1805.00048}{{\tt 1805.00048}}].

\bibitem{Blencowe:1988gj}
M.~P. Blencowe, {\it {A Consistent Interacting Massless Higher Spin Field
  Theory in $D$ = (2+1)}},  {\em Class. Quant. Grav.} {\bf 6} (1989) 443.

\bibitem{Bergshoeff:1989ns}
E.~Bergshoeff, M.~P. Blencowe and K.~S. Stelle, {\it {Area Preserving
  Diffeomorphisms and Higher Spin Algebra}},  {\em Commun. Math. Phys.} {\bf
  128} (1990) 213.

\bibitem{Pope:1989vj}
C.~N. Pope and P.~K. Townsend, {\it {Conformal Higher Spin in
  (2+1)-dimensions}},  {\em Phys. Lett. B} {\bf 225} (1989) 245--250.

\bibitem{Fradkin:1989xt}
E.~S. Fradkin and V.~Y. Linetsky, {\it {A Superconformal Theory of Massless
  Higher Spin Fields in $D$ = (2+1)}},  {\em Mod. Phys. Lett. A} {\bf 4} (1989)
  731.

\bibitem{Campoleoni:2010zq}
A.~Campoleoni, S.~Fredenhagen, S.~Pfenninger and S.~Theisen, {\it {Asymptotic
  symmetries of three-dimensional gravity coupled to higher-spin fields}},
  {\em JHEP} {\bf 11} (2010) 007 [\href{http://arXiv.org/abs/1008.4744}{{\tt
  1008.4744}}].

\bibitem{Henneaux:2010xg}
M.~Henneaux and S.-J. Rey, {\it {Nonlinear $W_{infinity}$ as Asymptotic
  Symmetry of Three-Dimensional Higher Spin Anti-de Sitter Gravity}},  {\em
  JHEP} {\bf 12} (2010) 007 [\href{http://arXiv.org/abs/1008.4579}{{\tt
  1008.4579}}].

\bibitem{Gaberdiel:2010pz}
M.~R. Gaberdiel and R.~Gopakumar, {\it {An AdS$_{3}$ Dual for Minimal Model
  CFTs}},  {\em Phys. Rev. D} {\bf 83} (2011) 066007
  [\href{http://arXiv.org/abs/1011.2986}{{\tt 1011.2986}}].

\bibitem{Gaberdiel:2012uj}
M.~R. Gaberdiel and R.~Gopakumar, {\it {Minimal Model Holography}},  {\em J.
  Phys. A} {\bf 46} (2013) 214002 [\href{http://arXiv.org/abs/1207.6697}{{\tt
  1207.6697}}].

\bibitem{Gaberdiel:2014cha}
M.~R. Gaberdiel and R.~Gopakumar, {\it {Higher Spins \& Strings}},  {\em JHEP}
  {\bf 11} (2014) 044 [\href{http://arXiv.org/abs/1406.6103}{{\tt 1406.6103}}].

\bibitem{Grigoriev:2019xmp}
M.~Grigoriev, I.~Lovrekovic and E.~Skvortsov, {\it {New Conformal Higher Spin
  Gravities in $3d$}},  {\em JHEP} {\bf 01} (2020) 059
  [\href{http://arXiv.org/abs/1909.13305}{{\tt 1909.13305}}].

\bibitem{Grigoriev:2020lzu}
M.~Grigoriev, K.~Mkrtchyan and E.~Skvortsov, {\it {Matter-free higher spin
  gravities in 3D: Partially-massless fields and general structure}},  {\em
  Phys. Rev. D} {\bf 102} (2020), no.~6 066003
  [\href{http://arXiv.org/abs/2005.05931}{{\tt 2005.05931}}].

\bibitem{Metsaev:2022yvb}
R.~R. Metsaev, {\it {Interacting massive and massless arbitrary spin fields in
  4d flat space}},  {\em Nucl. Phys. B} {\bf 984} (2022) 115978
  [\href{http://arXiv.org/abs/2206.13268}{{\tt 2206.13268}}].

\bibitem{Bekaert:2015tva}
X.~Bekaert, J.~Erdmenger, D.~Ponomarev and C.~Sleight, {\it {Quartic AdS
  Interactions in Higher-Spin Gravity from Conformal Field Theory}},  {\em
  JHEP} {\bf 11} (2015) 149 [\href{http://arXiv.org/abs/1508.04292}{{\tt
  1508.04292}}].

\bibitem{Steinacker:2020xph}
H.~C. Steinacker, {\it {Higher-spin gravity and torsion on quantized space-time
  in matrix models}},  {\em JHEP} {\bf 04} (2020) 111
  [\href{http://arXiv.org/abs/2002.02742}{{\tt 2002.02742}}].

\bibitem{Steinacker:2024unq}
H.~C. Steinacker, {\em {Quantum Geometry, Matrix Theory, and Gravity}}.
\newblock Cambridge University Press, 4, 2024.

\bibitem{Steinacker:2010rh}
H.~Steinacker, {\it {Emergent Geometry and Gravity from Matrix Models: an
  Introduction}},  {\em Class. Quant. Grav.} {\bf 27} (2010) 133001
  [\href{http://arXiv.org/abs/1003.4134}{{\tt 1003.4134}}].

\bibitem{Taylor:1999gq}
W.~Taylor and M.~Van~Raamsdonk, {\it {Multiple D0-branes in weakly curved
  backgrounds}},  {\em Nucl. Phys. B} {\bf 558} (1999) 63--95
  [\href{http://arXiv.org/abs/hep-th/9904095}{{\tt hep-th/9904095}}].

\end{thebibliography}\endgroup
\bibliographystyle{JHEP-2}

\end{document}